\documentclass[journal]{IEEEtran}

\ifCLASSINFOpdf

\else

\fi

\usepackage{rotating}

\usepackage{longtable}
\usepackage{lscape}
\usepackage{booktabs}
\usepackage{lscape}
\usepackage{colortbl}
\usepackage{multirow}
\usepackage{amssymb}

\usepackage{longtable} 
\usepackage{pdflscape} 
\usepackage{multirow}

\usepackage[english]{babel}
\usepackage[utf8]{inputenc}
\usepackage{amsmath}
\usepackage{graphicx}
\usepackage[colorinlistoftodos]{todonotes}
\usepackage{silence}
\usepackage{hyperref}
\usepackage{url}

\usepackage{breakurl}
\usepackage{siunitx} 
\usepackage{booktabs} 

\usepackage{graphics}
\usepackage{adjustbox}
\usepackage{amsfonts}
\usepackage{rotating}

\usepackage{longtable}
\usepackage{lscape}
\usepackage{booktabs}
\usepackage{lscape}
\usepackage{colortbl}
\usepackage{multirow}
\usepackage[T1]{fontenc}
\usepackage{multirow, makecell}
\usepackage{longtable}
\usepackage{subfig}

\newcolumntype{L}{>{\raggedright\arraybackslash}p}

\usepackage{mathtools}

\DeclarePairedDelimiter\abs{\lvert}{\rvert}%

\makeatletter
\let\oldabs\abs
\def\abs{\@ifstar{\oldabs}{\oldabs*}}
\usepackage{mathtools}

\newcommand\wfscore{$\textit{wF}1$}
\newcommand\wpscore{$\textit{wP}$}
\newcommand\wrscore{$\textit{wR}$}

\newcommand\mfscore{$\textit{mF}1$}
\newcommand\mpscore{$\textit{mP}$}
\newcommand\mrscore{$\textit{mR}$}

\newcommand\earlycrow{{\textsc{EarlyCrow}}}

\newcommand\pairflow{{\textsc{PairFlow}}} 
\newcommand\contextualsummary{{\textsc{ContextualSummary}}} \newcommand\made{{\textsc{Made}}} 
\newcommand\cc{C\&C}

\newcommand\dns[1]{{\small \texttt{#1}}}
\newcommand\dlabel\dns
\newcommand\shell\dns

\usepackage{tikz}
\newcommand*\circled[1]{\tikz[baseline=(char.base)]{
            \node[shape=circle,text=white,fill=black,draw,inner sep=0.5pt] (char) {\footnotesize#1};}}
            
\usepackage[ruled,vlined]{algorithm2e}

\usepackage{multirow, makecell}

\newcommand{\independent}{\perp\mkern-9.5mu\perp}

\usepackage{lipsum}

\usepackage{listings}
\usepackage{xcolor}
\begin{document}

%
\title{Detecting APT Malware Command and Control over HTTP(S) Using Contextual Summaries}
%
%
%

\author{Almuthanna~Alageel and Sergio~Maffeis \\
       \IEEEauthorblockA{Department of Computing} \\
              \IEEEauthorblockA{Imperial College London} \\
       \IEEEauthorblockA{London, United Kingdom} \\

}

%
%




\thispagestyle{plain}
\pagestyle{plain}
\maketitle

\begin{abstract}

Advanced Persistent Threats (APTs) are among the most sophisticated threats facing critical organizations worldwide. 
APTs employ specific tactics, techniques, and procedures (TTPs) which make them difficult to detect in comparison to frequent and aggressive attacks. 
In fact, current network intrusion detection systems struggle to detect APTs communications, allowing such threats to persist unnoticed on victims' machines for months or even years. 

In this paper, we present \earlycrow, an approach to detect APT malware command and control over HTTP(S) using \emph{contextual summaries}.
The design of \earlycrow\ is informed by a novel threat model focused on TTPs present in traffic generated by tools recently used as part of APT campaigns.
The threat model highlights the importance of the context around the malicious connections, and suggests traffic attributes which help APT detection. 
\earlycrow\  defines a novel multipurpose network flow format called \pairflow, which 
is leveraged to build the contextual summary of a PCAP capture, representing key behavioral, statistical and protocol information relevant to APT TTPs.
We evaluate the effectiveness of \earlycrow\ on unseen APTs obtaining a headline macro average F1-score of 93.02\% with FPR of $0.74\%$ \footnote{This paper is an extended version of Alageel, Almuthanna, and Sergio Maffeis. "EarlyCrow: Detecting APT Malware
Command and Control over HTTP(S) Using Contextual Summaries." In the 25th International Conference, ISC 2022, December 18–22, 2022, Proceedings, pp. 290-316. Cham: Springer International Publishing, 2022.}.  
\end{abstract}


%
\IEEEpeerreviewmaketitle

\section{Introduction}

\IEEEPARstart{A}{dvanced}, Persistent Threats (APTs) are known to be the most sophisticated long-term attack campaigns targeting highly protective organizations \cite{ahmad2019strategically}. 
APTs are generally aware of internal defenses related to their target \cite{Schindler2017AnomalyDI}, and usually do not send spam, participate in DDoS attacks, or aggressively propagate to other hosts to spread infections at scale \cite{farinholt2020dark}. 

APT malware are those malicious tools known to be used by APT campaigns. 
The most common is the Remote Access Trojan (RAT), typically composed of a builder, stub, and controller. The builder initiates a new instance stub upon the infection. The stub runs on the victim machine and contains a hard-coded Fully Qualified Domain Name (FQDN) or IP to communicate to the RAT controller, which resides on the Command and Control (\cc) server \cite{rezaeirad2018schrodinger}. 
Rootkits, spyware, downloaders, and keyloggers may also be part of an APT campaign. 
APT malware such as DarkComet includes these functions in one ecosystem  \cite{farinholt2020dark}, which may capture the audio, explore files, and drop malicious tools through visiting URLs \cite{farinholt2017catch}. Griffon, used by FIN 7, can gather information, load Meterpreter, and take screenshots \cite{griffon19}. 
Hutchins et al. \cite{hutchins2011intelligence} propose a kill chain to defend against APTs at various stages, including reconnaissance, weaponization, delivery, exploitation, installation, and \cc.
These stages normally iterate over a long time \cite{milajerdi2018holmes}. 
In order to limit the damage inflicted by an APT, it is essential to detect them at an early stage, and in particular as they establish communication with the \cc. 
By inspecting honeypot data, we find that the communication to \cc\ starts immediately once the machine is infected. Several automated tasks are performed, including establishing \textit{fallback channels} and downloading further payloads from the \cc\ server. These activities intentionally behave as legitimate web browser activities, attempting to evade Network Intrusion Detection Systems (NIDSs).

Typical botnet behaviour includes launching DDoS, propagating quickly to other hosts, internal scanning, or sending spam~\cite{gu2008botminer}. In the past, anomaly-based and NetFlow feature-based were able to detect such behaviour and filter it out, detecting the abnormality of a connection over a relatively short period of time. 

APT operators deploy several TTPs to keep a low profile and facilitate incoming targeted and stealthy attacks over a long period. Therefore, relying on the traffic frequency, beaconing behaviour, or NXDOMAIN generated by DGA in many botnets is insufficient to detect APT malware. 
Also, the existing defences that adopt feature-based machine learning or deep learning \cite{bilge2012disclosure,yen2013beehive,hu2016baywatch,oprea2015detection,perdisci2010behavioral,van2018applying} to detect non-targeted attacks, are much less effective at APT detection, as we will discuss in Sections \ref{sec:EarlyCrow_threat_model}, \ref{sec:earlycrow_desc} and \ref{sec:EarlyCrowevaluation}. 
Deep learning requires a vast dataset to train a classifier, but real APT data is scarce. Machine learning with feature extraction is a promising candidate, but we need first to identify the tailored features that will work specifically for APTs, as proposed in \cite{alageel2021hawkeye} for DNS infrastructure.

The quote says "Information is power only if you can take action with it" \cite{burrus2018q}. All network data is recorded in PCAP format, which is difficult to follow and store in practice. A number of tools have been proposed to extract the desired fields and specific context for network and security operations. The state of the art in processing PCAPs to produce logs for detection and forensics purposes is not adequate to fully capture APTs. Using NetFlow or Web Proxy logs miss several essential fields, which limit APT-related feature extraction. In addition, considering the sparse logs files generated by tools such as Zeek/Bro\cite{paxson1999bro} causes overhead to understanding the contextual behaviour of their file formats. Apart from raw PCAP, these data formats do not provide a way to access the packet-level data points, including bytes, timestamps, and the protocol, which hinders the researcher from providing important features relevant to APTs. For instance, we notice that APTs connections have considerably larger mean delta packets interarrival times. Likewise, the data packet exchange idle time of APTs is significantly different from legitimate traffic. Nevertheless, these features do not suffice to clearly separate the classes of interest, as some legitimate connections, browsing-like, have similar behaviour. 

That leads to further investigation of other attributes, such as the number of failed HTTP packets, which is more prominent in APTs. Hosts may use several User-Agents, some of which are already reported as an IoC for malware using the exact string, which may conflict with the list of UA's a host. In addition, the list of FQDNs, Resource Records, and URLs accessed by users is vital to be tracked. The type of client cipher suits used by the host through a specific time may reveal a malicious behaviour of an APT malware pre-configured to be using particular settings. We discuss the issue of the current data format further Section~\ref{sec:data_format}.

In this paper, we propose a novel threat model that highlights the importance of the context around the malicious connections, and suggests traffic attributes of TTPs which help APT detection (Section~\ref{sec:EarlyCrow_threat_model}). We focus on two cases in our threat model and introduce the first public measurement study of malware used by APTs and botnet \cc\ communication to investigate the deployed TTPs.
We leverage our measurements to identify the features necessary to recognize such TTPs at the network level, and compare them with existing features from the literature. We found that the use of evasive TTPs leads to significant overlap with legitimate behaviour, confusing the decision boundaries based on known features (Section~\ref{sec:measurements}). 

Based on this analysis, we build \earlycrow, a tool to detect APT activity in network traffic. \earlycrow\ utilises  \pairflow, a novel multipurpose network flow format which captures the necessary fields of the connections between host pairs of interest, over a granular time tuned by security practitioners (Section~\ref{sec:cf_description}).
\earlycrow\ generates four sets of data focused on connections, hosts, destinations, and URLs. Features from these sets are grouped to form a \contextualsummary. The \contextualsummary\ has multidimensional features that help in building more informative random forest trees used for classification as described in Section \ref{sec:earlycrow_desc}.

We evaluate \earlycrow\ on traffic from APT-like malware excluded from the measurement study and training set in order to test generalization and mimic a real-world scenario (Section \ref{sec:EarlyCrowevaluation}). Fresh malware samples are also investigated to confirm the feature importance identified by our measurement study on the training set.
We also investigate how the performance of \earlycrow\ is affected by different deployment scenarios, where it has visibility on HTTP traffic or where it can only observe opaque HTTPS traffic. 
\textsc{EarlyCrow} defends well against unseen APTs that encrypt their traffic with HTTPS, obtaining a headline \mfscore of 93.72\%, and accuracy of 98.11\% with FPR of 0.74\%. In comparison, the state of the art stands at 60.29\% and with no false positive rates.

In summary, our main contributions\footnote{For IEEE reviewers: this manuscript includes new material not included in the previous conference paper. We provide an expanded analysis of the evidence-based threat model in Section \ref{sec:EarlyCrow_threat_model}, highlighting the shortcomings of the current data format exploited by APTs. Section \ref{sec:cf_description} explores the technical details of \pairflow, which serves as a critical foundation for \earlycrow\ and APT detection in general. In Section \ref{sec:datasets}, we describe our collected dataset and summarize the corresponding campaigns associated with specific attacks. Causal analysis is employed to troubleshoot and audit our datasets, as detailed in Section \ref{appendix:causal}. Furthermore, we conduct additional experiments in Sections \ref{sec:EarlyCrowevaluation_known} and \ref{sec:EarlyCrowevaluation_unseen}, reporting weighted and micro metrics alongside further discussions and findings. Finally, in Section \ref{sec:featuresTTPsCorr}, we present additional experiments to demonstrate the correlation between TTPs and the proposed features.} are:
\begin{list}{$\bullet$}{\leftmargin=1em \itemindent=0em}
\setlength{\itemindent}{0.1em}
    
    \item We present an evidence-based analysis of various TTPs used by APTs. These TTPs are known to be used to evade NIDS \cite{MitreCnC}. We also introduce a measurement study on various APT malware over popular and novel features to capture TTPs usage.

    \item We implement \earlycrow \footnote{\earlycrow\ code, datasets, and experiments are publicly available at \cite{EarlyCrowRepo}.}, a tool to detect evasive malicious communication over HTTP(S). \earlycrow\ focuses primarily on APTs but is also effective against stealthy botnets.
    
    \item We evaluate the classification performance of new and existing features for malicious traffic detection under different scenarios distinguishing ATP, botnet, and legitimate traffic.

\end{list}


\section{Threat Model}\label{sec:EarlyCrow_threat_model}
\begin{figure}[!t]
\centering
\includegraphics[width=9cm,height=14cm]{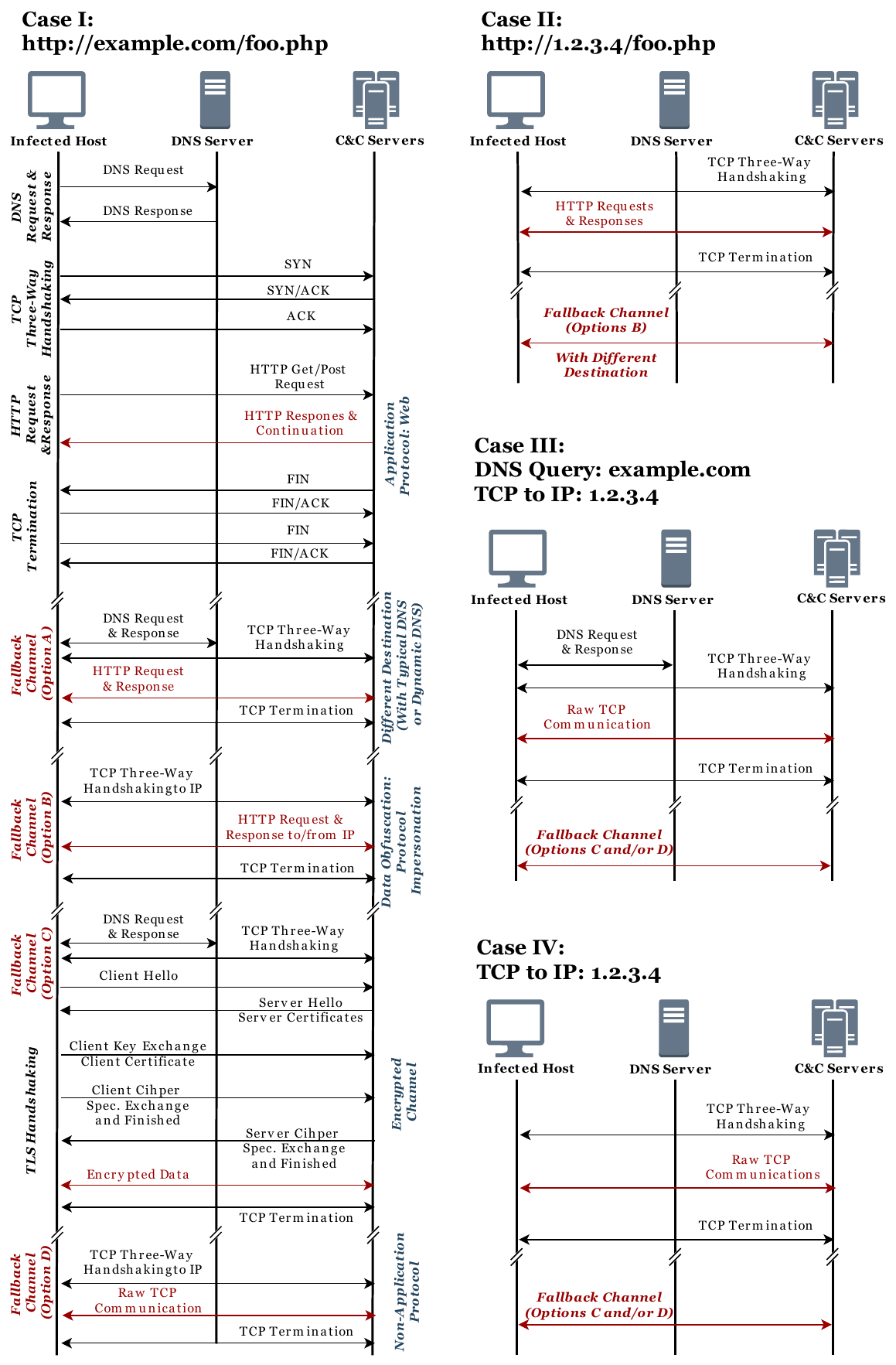}
\caption{Types of APT malware initial communications. Fallback channels may use one or more TTPs (blue text). Malicious data packets are represented by red arrows.}\label{fig:traffic_threat_model}
\end{figure}

Defining a relevant threat model and focusing on a narrow set of attacks are recommended best practices when proposing a novel NIDS \cite{sommer2010outside}.
There are several ways to analyze the threat model for APTs at the network level. While \cite{alageel2021hawkeye} focuses on the DNS infrastructure, this paper focuses on APTs traffic. We identify four popular cases, each of which may deploy at least one \cc\ TTP, as depicted in Figure \ref{fig:traffic_threat_model}. 

In Case I, the infected machine contains APT malware with a hard-coded FQDN. The malware issues a DNS query to resolve the FQDN to an IP address. The subsequent communication to the \cc\ server can be via HTTP or HTTPS. After that, the malware may initiate a \emph{fallback channel}, another popular TTP used by APTs \cite{TTPFallback}, using either of the strategies described in Cases I-IV, only this time no longer for the initial communication. 
In Case II, the APT malware connects to a URL whose domain component is a hard-coded IP address, in order to bypass malicious domain detectors, and its fallback channel can be established using the \emph{DNS over HTTPS} (DoH) TTP \cite{TTPDoH}, as in  CobaltStrike \cite{JansenCobaltStrike}, which is used by SUNBURST \cite{SUNBURSTFireEye}.
Case III is similar to Case I in using a hard-coded FQDN, but the subsequent communication uses raw TCP rather than HTTP during the malicious operation. 
Case IV is similar to Case II in using direct IP without DNS resolution, but then uses raw TCP communication as in Case III. Both Case III and IV may use fallback channels with various TTPs, although not including those related to HTTP(S). 

Additional TTPs introduced by MITRE and relevant to APTs can be combined with the use of a fallback channel: %
\emph{web protocol} \cite{TTPweb} where an adversary may use HTTP to avoid network filtering and mimic legitimate and expected connections, 
\emph{non-application protocols} \cite{TTPnonApp} such as Raw TCP, UDP or ICMP, 
\emph{encrypted channel} \cite{TTPencryptedChannel} to hide \cc\ malicious content, \emph{fast flux} \cite{TTPFastFlux} (a sub-technique of \emph{dynamic resolution}) to obtain different IPs for the same FQDN,
and \emph{data obfuscation through protocol impersonation} \cite{TTPimppersonation} to impersonate legitimate use of HTTP or to mimic a trustworthy entity using a fake SSL/TLS certificate. 

This paper focuses on Cases I and II, where at least one malicious HTTP(S) connection exists between the infected host and \cc\ server. Cases III and IV are left for future work.

\subsection{Current Data Formats Issues}\label{sec:data_format}
PCAP is the de-facto format for network captures, and it provides the raw data at the network level. However, several NIDSs use different data formats built on top of PCAP to improve detection performance. Current data formats hinder security practitioners from extracting specific fields to capture the IoC, such as URLs and IP addresses. Many informative fields have to be summarised for the end-to-end connections to provide the contextualisation to extract instructive features that reduce FPR. Examples of these fields include A and nameservers resource records, HTTP status codes, content types, and protocols used during the connection.  

\textit{NetFlow} is a popular data format, which is defined in RFC 3954 \cite{claise2004cisco}. It provides the network operator with IP flow information on their network for diagnosis and troubleshooting purposes \cite{claise2004cisco}. Several packets between a source and destination with common properties, such as protocol and port, are encapsulated in one flow for a specific time window. 
%
Security analysts and researchers widely use NetFlow to detect malicious activities. However, NetFlow does not provide salient information to a flow, such as FQDN, accessed URL, UA, SSL/TLS settings, and HTTP status codes counts. In addition, a flow cannot include more than one protocol, which may increase the challenge for security analysts and NIDS to detect long, low-mode malicious activities.

\textit{Tranalyzer} \cite{burschka2016tranalyzer} is another lightweight flow generator and packet analyzer developed for researchers and practitioners. It enhances the capabilities of NetFlow and aids analysts in analyzing massive packet dumps. It also provides access to drill down to the specific flows or even individual packets. However, Tranalyzer separates flows according to their protocol, which does not provide contextual data between pairs of endpoints. %

\textit{Contextual Flow} \cite{anderson2016identifying} is another flow-based format designed to detect encrypted malicious traffic without decryption. The contextualization attaches the TLS parameters and UA to the flow. 
However, Contextual Flow lacks accessing the details of packet arrival time that forms such a flow, which may weaken the ability to analyze stealthy attacks during a flow. Therefore, HTTP responses behaviour cannot be analyzed. FQDN resource records and domain age are not included in a flow, requiring other data formats for malicious domain detection. The contextualization is also lacking from tracking the previous flow, which can help to capture those attacks successfully and evade NIDS, but can be reconstructed over time for detection.

Other popular data formats provide part of the required fields to capture APT attacks. \textit{Passive DNS} was introduced by F. Weimer \cite{weimer2005passive} to detect malicious connections resolved by DNS. Its output captures DNS responses and extracts resource records without involving authoritative name servers. However, relying on DNS resource records and their TTL is insufficient to detect malicious connections, including APTs.
\textit{Web Proxy} is also a popular data format for parsing HTTP requests and responses and log-related information in a tabular format. Examples of web proxy popular generators include Microsoft Forefront TMG \cite{forefrontWebProxy}, and Squid \cite{SquidWebProxy}. APTs may use HTTP(S), raw TCP, or other protocols. Therefore, web proxy logs may not be the best to reconstruct APT attacks.

\textit{Zeek/Bro} \cite{paxson1999bro} is a network intrusion detection system run in real-time. However, several systems use Bro to compile PCAP files and generate more than 20 data formats to be used as input for proposed systems. These logs include connection status, application statistics, DNS, DHCP, HTTP, SSH, etc. However, Bro does not generate a flow-based format that tracks the story of a given connection over a period of time; instead, it provides a sparse format that needs considerable time to extract data related to a connection. 
%

\subsection{TTP Relevant Data}
In order to define a convenient data format which can be used by feature-based NIDS for APTs communications detection, it is essential to understand the threat hunting of APTs in practice.
To track TTPs at the network level, a security researcher needs to collect `static' IoCs for known APTs or manually identify the likelihood of specific traffic-based TTPs based on enterprise network behaviour. Security vendors publish IoCs of discovered APTs, which may present a usage of TTPs like DGA or Dynamic DNS. These IoCs help security engineers to investigate if their enterprise is compromised by an APT campaign. Another source of information beyond IoC is the industry reports. These reports can explain the traffic behaviour of the infected enterprise. For example, an APT campaign may frequently use a non-application protocol or a fallback channel, which is unusual traffic for the infected enterprise. 

In this section, we present the two methods, IoC-like and traffic data, to study the APTs based on such information, which is presented in Section \ref{sec:measurements}. This is also important when we propose our data format in Section \ref{sec:cf_description}.

\subsubsection{IoC-like Data}
The proposed data format should provide enough information to extract IoCs from PCAPs. These include FQDN, IP, URL, UA, and Encryption Settings.
%
APT campaigns attempt to reserve one or more \textit{FQDN(s)} to locate \cc\ servers. They may mimic the targeted organisation interest or use Dynamic DNS, which is another TTP\cite{TTPdynamicDNS}, to communicate back to \cc\ servers~\cite{farinholt2020dark,alageel2021hawkeye}. The resolved FQDN holds at least one A resource record and many records for popular web servers. However, some APTs provide many A resource records to provide a fallback channel for the next sequence connection. 

\textit{URL} is known to be used as an IoC and used in HTTP-based malware detection \cite{ma2009beyond, oprea2018made, perdisci2010behavioral,bartos2016optimized}.
Some APT malware downloads an executable file or passes another malicious FQDN, IP, or configuration commands in the new URL parameters in subsequent requests. A typical URL structure includes FQDN, nested folders (which we will referr to as \emph{depth}), filename, parameters and values with a delimiter (\&) to separate between them and (=) to assign value to the parameter, and encoded strings which typically contain \%-encoding.

\textit{User-Agent (UA)} is also another IoC used in malicious traffic detection \cite{kheir2013behavioral, oprea2018made, anderson2016identifying}.
Different browsers are available to surf websites, and they vary in supporting technologies such as frames, images, and video content. Web developers design websites to fit the majority of browser rendering abilities. For malware, it is common to find typos, outdated versions, or even leaking information \cite{wang2016trafficav}. Other malware has been found to use an empty string (e.g., Trojan.dropper), unknown string (e.g., H-worm, Win32/Sality.3, Win32.QQP), or inconsistent string (Win32/Alman, Trojan.Win32.Dropper.aa) \cite{grill2014malware,anderson2016identifying}. %
%
%
%

The last IoC to check is the \textit{Encryption Setting}.
Some APT malware uses malicious TLS certificates to impersonate trusted entities (\textit{data obfuscation: protocol impersonation TTP}), or to hide the traffic payload from an IDS. It is vital to check cipher suit settings to find whether the adopted one is insecure, unusual, or weak, which is a common sign of malware \cite{csinfo}.

\subsubsection{Traffic Data}
Although multiple traffic-based TTPs were used by APTs in the past, it is challenging to capture them by configuring NIDS with naive and straightforward rules, which may raise the FPR sharply. Hence, we need to consider how malicious packets are sent at the contextual level. First, we need to cover the details of HTTP requests and responses, and then the traffic behaviour of all protocols used for the same flow.
\textit{HTTP request and response context} involves consecutive HTTP transactions composed of request and responses. A request is mainly characterized by the URL, method type (e.g. GET, POST .. etc), UA and Referrer. Response headers specify the settings of the content type, Cookie and status codes. To detect APT malware, we need to efficiently store that information between two endpoints in one flow and enable the NIDS to extract valuable statistics at the packet level. 
%

Due to the stealthiness and low-mode operation for APTs, we also need to provide a way to investigate \textit{Traffic behaviour}. This can be achieved by storing the packets' arrival time, their length and other related information presented in Section \ref{sec:cf_description}.  
Such a summary needs to cover the control and data planes of TCP,  UDP and ICMP packets. With this summary on datapoints, NIDS designers can catch APTs TTPs such as \textit{fallback channel} and using \textit{non-application protocols}. For instance, a host contacting three different destinations after only one DNS query can be a sign of infection by the fallback channel technique. Another scenario of TTP, the \textit{non-application protocol}, is when the APT malware opens a legitimate-looking HTTP connection but that is followed by a sequence of malicious raw TCP packets as depicted in Figure \ref{fig:traffic_threat_model}.

\section{Traffic Measurements}\label{sec:measurements}
We provide several measurements taken on the training set summarised in Table \ref{tab:EarlyCrowdataset}, page \pageref{tab:EarlyCrowdataset}, and described in Section~\ref{sec:datasets}. Since our objective is to detect APTs at the early stage, all measurements are observed during the first 15 minutes of each connection, which is the longest duration allowed by the Any.Run sandbox\footnote{https://any.run}. In Section \ref{sec:EarlyCrowevaluation_unseen}, we will investigate these measurements and other proposed features, to see if they generalize to unseen malware. 

\subsection{Traffic Statistical Measurements}
\begin{figure*}[!t]
\centering
\includegraphics[width=15cm,height=19cm]{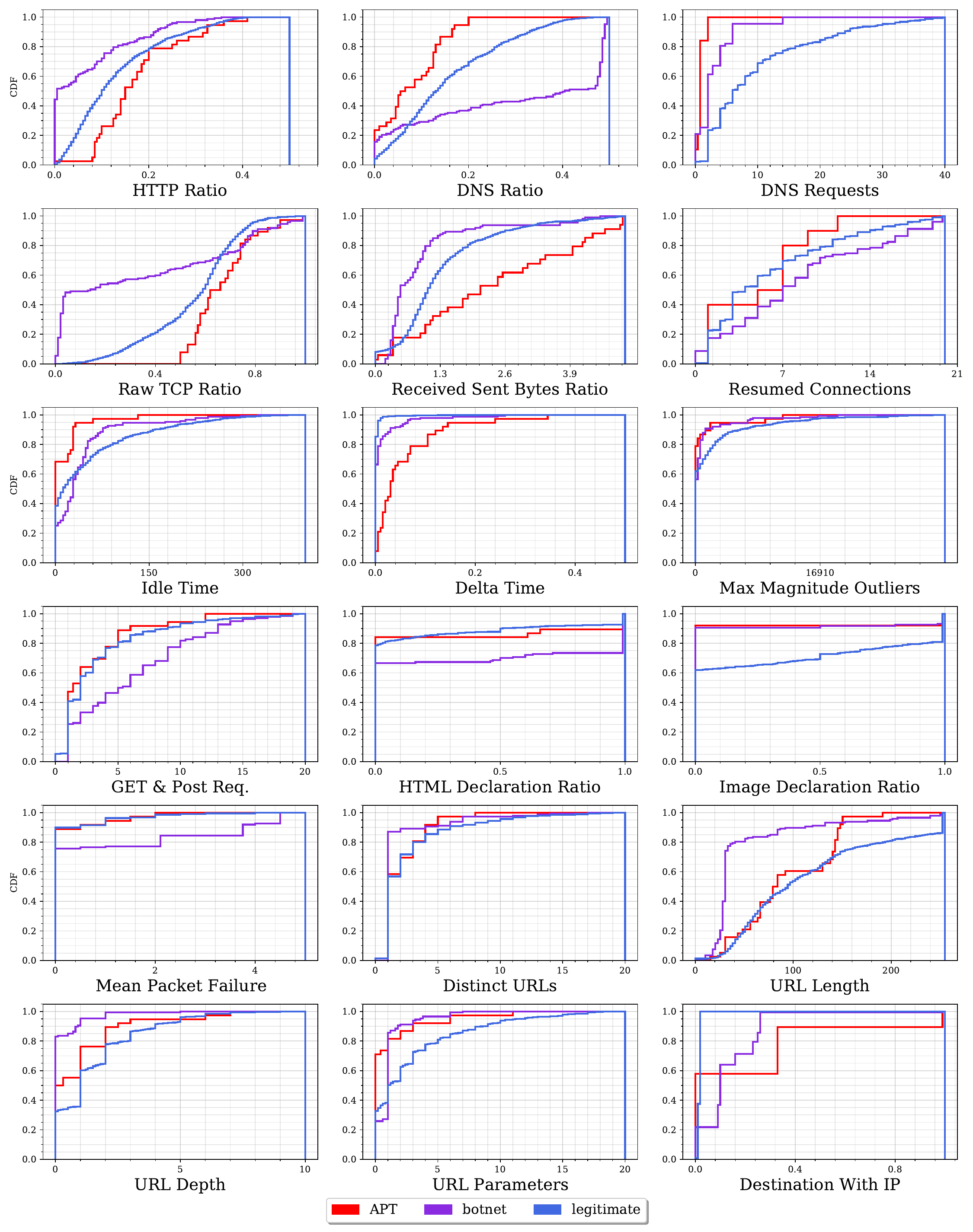}
\caption{Measurements for APT, botnets, and legitimate connections.} \label{fig:measurements}
\end{figure*}

Statistical end-to-end observations may highlight the evasive behaviour of APTs compared to legitimate actors. The presence of a slight deviation may reflect malicious use of three TTPs, including \textit{non-application protocols}, \textit{data obfuscation through protocol impersonation} and \textit{web protocol}. 
Since this study focuses on malicious HTTP(S) usage, we measure the HTTP packets ratio across all classes. Other related protocols are also measured, including raw TCP and DNS ratios. 
Legitimate connections show a positive linear relationship between DNS and HTTP packets (Figure \ref{fig:measurements}). 
With every additional page requested by a user, such packets are exchanged with a remote web server in order to fetch additional resources. 
For APTs, we notice that DNS ratios are half or less than for legitimate or botnets, respectively. 95.2\% of APTs do not exceed a 0.19 DNS ratio, compared to 0.38 for legitimate and 0.46 for botnets. 

Next, we focus on DNS requests and conclude that almost no malicious behaviour exceeds the legitimate, except for Conficker botnets, which use the DGA technique. 84\% of APT or botnet traffic issues 2 or 6 requests at most, while legitimate traffic can generate up to 18. 
Once a domain is resolved to one or more IPs, a typical APT avoids requesting another DNS for the rest of HTTP communication unless they plan to establish another \emph{fallback channel}.
Another useful feature is the raw TCP ratio, which helps detect the \textit{non-application protocols} TTP: a high ratio indicates the adversary uses HTTP as camouflage while still heavily relying on raw TCP.  
It is extremely rare for an APT to have a raw TCP ratio lower than 48.84\%, whereas we observed minimum ratios of 2\% of legitimate, and 0\% of botnets.

Since we focus on the early stage of connections originating from the victim side, we found that around 70.58\% of APTs receive 3.35 times more data than they send to the remote server, compared to 1.45 and 0.75 for legitimate and botnets, respectively. 
This is consistent with the threat model in \cite{alageel2021hawkeye}, where an adversary uploads more tools on the victim's machine at the beginning of an APT campaign to continue other operations such as lateral movement, unlike botnets, which may show more data exfiltration behaviour. 
%
We also examine the number of resumed connections. Legitimate HTTP usage typically increases the number of resumed connections. Once the web resources are downloaded and the keep-alive time has expired, the TCP connection is terminated with FIN.  
Upon clicking another link, even for the same website, a new TCP three-way handshake is initiated.
We count that as a resumed connection.
With a web caching service, the scenario remains similar, although the server is contacted via a proxy or content delivery network (CDN). 
While legitimate and botnets connections may easily be resumed up to 21 times, APTs tend to terminate less (roughly 50\% less). %
It seems plausible that APTs avoid frequent connection termination and resumption to increase stealthiness.

\subsection{Time-based Measurements}
We measure \emph{stealthiness and low profile} of APTs by monitoring time-based features. 
\emph{Delta}, the packet inter-arrival time between a remote server and a host, is estimated based on the arrival time difference between packets, independently from their protocol. 
For 94.73\% of cases, we found the mean delta time in seconds to be at most $23.5\times 10^{-2}$ for APTs, $6\times 10^{-2}$ for botnets and $ 0.5\times 10^{-2}$ for legitimate. Hence, APTs may act slower than botnets and legitimate by up to 4 and 47 times respectively. 

A new metric, data packet exchange \emph{idle time}, is proposed to measure the time difference between actual data packets. 
We found APTs idle time to be 3 and 6.57 times shorter than botnets and legitimate: 92\% of cases have an idle time of at most 28, 84, and 184 seconds, respectively.
%
%
Once APTs establish a communication  channel, they send bursts of data packets (low idle time), then pause communications (high delta) until the next burst. 
We also measure the maximum magnitude of outliers which exceed the Simple Moving Average (SMA) with respect to the predefined bins described in Section \ref{subsec:time_based}. We found that for 84\% of the cases, the maximum magnitude for APTs (0.338 KB) is half the one for botnets (0.676 KB), and 10\% of the one for legitimate (3.33 KB).
These three time-based features partially capture the \emph{low and stealthy profile} of APTs compared to botnets or legitimate.

\subsection{Remote Web Server}
Analysis of contacted web servers may help identifying the \textit{web protocol} and \textit{fallback channel} TTPs.
Typical web servers mostly adhere to best practices in setting up their HTTP configurations. 
%
%
APTs appear to be more professionally configured than botnets, but not as much as legitimate ones. 
For instance, the \emph{packet failure} rate for legitimate servers and APTs (HTTP responses with status codes 4xx and 5xx) is relatively low. To be precise, 90\% have at most one packet failure, while the botnets may receive as many as five. 
Total GET and POST requests are less similar. 92\% of APTs and legitimates have 9 and 10 or less, respectively while the botnets have up to 14.
%

We also investigate the ratios of content types declarations.
We focus on the ratios of HTML and images, since these are most frequently used in HTTP connections. 
73\% of APTs, legitimate and botnets declare HTML 2\%, 2\% and 98\% of the time, so APT behaviour, in this case, is similar to legitimate.
However, due to the possible use of the \textit{data obfuscation through protocol impersonation} TTP, we found that APTs and botnets are less likely to declare image type, which is not the case for web browsing activities: 
70\% of legitimate declare images 30\% at most during a connection, while it is zero for both APTs and botnets. 
%

Next, we measure the URL characteristics, due to their proven effectiveness in detecting malicious web servers. 
Measuring the distinct URLs accessed in a given network may highlight the rich number of web pages which is more likely to be legitimate \cite{oprea2018made,hu2016baywatch}. We observe that APTs invest heavily in legitimate-looking pages, to evade NIDS that rely on URL-based features. 
For example, we find that 87\% of botnets query only one URL, while legitimate and APTs query up to five and four, respectively. 
APTs have more resources than botnets in general. 
As depicted in Figure \ref{fig:measurements}, 90\% of APTs have 3 nested folders (depth), close to legitimate, which is 4, while botnets have 1 at most. URL parameters differ even more: 87\% of APTs and legitimate use 3 and 7, while botnets use only 1. 
Following that, URL length is determined by the length of FQDNs, depths, filenames, parameters, values, fragments, and strings. 90\% of legitimate URL lengths are 249 or less, whereas APTs and botnets are up to 145 and 109.
Finally, APTs deploy a fallback channel in several ways, as discussed in Section \ref{sec:EarlyCrow_threat_model}.
We measure the number of HTTP(S) connections established to an IP without a previous domain resolution. 
57.89\% of APTs reached 32\% of \cc\ with IP only, while it is 9\% and 1\% for botnets and legitimates.  
Therefore, it is unusual for legitimate to perform such behaviour, while it is more common for APTs and occasional for botnets.

In the next section, we will describe \pairflow, which allows the NIDS designer to quickly pivot flows into many profiles such as host, destination, URL profiles. It also allows being used by those malicious domains or IP detectors. Instead of detecting one flow according to their initiating and termination of TCP, protocol-based or time window, it digests all information to extract features later based on the whole context over time. So it is vital to collect all proper IoC and traffic-based TTPs known to be used for malicious detection.

\section{\pairflow} \label{sec:cf_description}

\pairflow\ is a data format that allows the NIDS designer to quickly pivot flows into many profiles such as host, destination, URL profiles. \pairflow\ data can also be used by detectors of malicious domains or IPs. Instead of detecting one flow according to the initiation and termination of TCP, protocol-based or time window, \pairflow\ digests all information to extract features later based on the whole context over time.

%
\subsection{Architecutre Overview}
\pairflow\ receives raw PCAP data and stores these packets in a buffer until a time window of size $t$ has passed. 
The buffer sends the current granular data of a time window of all connections at an enterprise network to the \emph{Tracking} module to group unique pairs and label related packets (Figure \ref{fig:pairflow_architecture}, \circled{1}). A unique pair refers to any (possibly bidirectional) connection observed between a host on the local network and a remote server. We take the \emph{source} of the pair to be the local host, and the \emph{destination} to be the remote server. 
Next, the \emph{Aggregator} module adds a \pairflow\ ID and time window (Figure \ref{fig:pairflow_architecture}, \circled{2}) to the flow data. 
The Aggregator module is also responsible for marking packets according to their plane, extracting the domains and HTTP fields. 
Next, the \emph{Encapsulation} module groups all these pieces of information contextually (Figure \ref{fig:pairflow_architecture}, \circled{3}), so that all possible TTPs in Figure \ref{fig:traffic_threat_model} can be analyzed later. Therefore, each pair of connections has a comprehensive description of their packets behaviour (described in Section \ref{subsubsec:packet_behavior}), HTTP settings, accessed domains, and cipher suites setting. Finally, \pairflow\ outputs four additional JSON files which can be used by any external classifier (Figure \ref{fig:pairflow_architecture}, \circled{4}).

\subsection{Tracking}

\subsubsection{Packets Retrieving.}
The tracking module identifies all unique pair connections on the network and filters out those using non-IP protocols. 
For each unique pair connection, \pairflow\ tracks, bidirectionally, all packets related to a pair.
These packets are designated with an initial Flow ID. 
The Flow ID holds unchanged for all packets during the same time window for a given pair connection. 
Each packet will maintain its individual index for the aggregation step later.
Packets with the same Flow ID may also use different protocols.
Therefore, each one has a one-hot encoding flag called Encoding Protocol Flag (EPFLAG) used later for further filtering. 
These flags started with EPFLAG\_Protocol, where a protocol is a subset of \{TCP, UDP, DNS, ICMP, HTTP, SSL/TLS\}.

\subsubsection{DNS Requests and Responses.}
The tracked packets do not include DNS requests and responses, which are responsible for locating the IP address needed to establish a connection. That is due to the pair connection being between the host and the DNS server, which is different than the destination. Similar to \cite{anderson2016identifying}, to track these DNS packets, a destination of the present pair will be used as a Local PTR to find all DNS response packets from the PCAP repository. Once found, the DNS response resource records will be used to find all related DNS requests. Now, Any packets belonging to the pair connection are attached and sorted according to their arrival time. Those packets outside of time window are not included.

\begin{figure*}[!t]
\centering
\includegraphics[width=18cm,height=10.5cm]{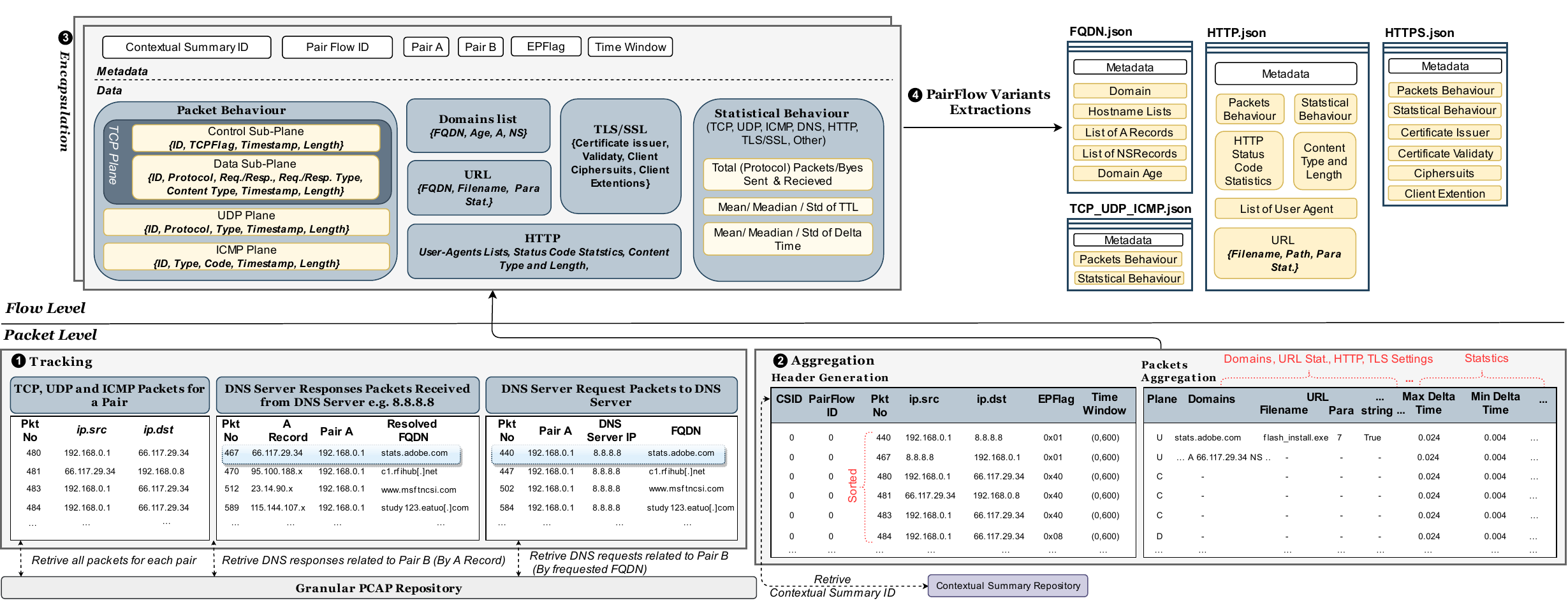}
\caption{Overview of the \pairflow\ workflow.}\label{fig:pairflow_architecture}
\end{figure*}

\subsection{Aggregation}

\subsubsection{Header Generation.} 
Besides the individual packet ID from the PCAP, every packet is also designated with a Flow ID composed of a \contextualsummary\ ID, described in Section \ref{sec:contextual_summary}, and a \pairflow\ ID.
The former is unique for the lifetime of a pair, while the latter is unique for a time window. 
Any packets from that \pairflow\ will always have the same Flow ID. 
To assign the PFID, the Aggregation module will check the \contextualsummary\ repository to find if the pair has been processed in the past. 
If so, the incoming PFID will be the last used PFID for the same pair and \contextualsummary\ ID, incremented by 1. 
Otherwise, a new and unique \contextualsummary\ will be created, and the PFID will start with 0.

\subsubsection{Packets Aggregation.}
The Aggregator module creates a \pairflow\ to store \pairflow\ ID, sorted packet index, pair connection, time window, EPFlag, FQDNs, URL, UAs, SSL/TLS settings, and initial flow-based statistics.
The initial flow-based statistics include the number of protocol-based packets (i.e., TCP, UDP, ICMP, HTTP, SSL/TLS, DNS packets), total (encrypted) bytes, and total (encrypted) bytes sent/received. 
Time-based statistics include packet Time to Live (TTL) and delta packets interarrival time max/min/median and the flow duration at the same time window. 
Similar to \cite{assadhan2008network}, we separate TCP packets into data and control packets to be used later in the encapsulation process. 
Finally, preprocessed flows are dispatched to the encapsulation step for further processing.

\subsection{Encapsulation} 
The encapsulation phase explicitly groups packet behaviour, FQDN and URL, HTTP(S) and initial statistical behaviour implicit in preprocessed flows, in order to make contextual information readily available. 
The data types involved include a list of strings and tuples, Boolean and numeric fields, as shown in Table \ref{table:cf_feature}. 

\subsubsection{Packet behaviour}\label{subsubsec:packet_behavior} 
Packet behaviour encapsulates all packets according to their protocol type (TCP, UDP, and ICMP) in a list of tuples. The first element is the packet index for traceability of a given packet inside the original PCAP for further investigation.

The \textit{TCP plane} involves the control and data sub-planes as shown in Figure \ref{fig:pairflow_architecture}. 
Each packet in the data sub-plane holds protocol name, request/response and their type, content type, timestamp, and packet length for each packet. For example, an HTTP request packet can be described as (460854, 'HTTP', 'Request', 'GET', 'Empty Content', 1066.51, 383) and its response (460895, 'HTTP', 'Response', 200, 'text/javascript', 1066.86, 429). This helps the upper system work on time series traffic and monitor the anomaly for a given \pairflow. Further packet-level statistical analysis such as counting GET/POST, HTTP Response types, content analysis can be achieved as described in Section~\ref{sec:earlycrow_pairflow_features}.
The control sub-plane provides the behaviour of the initial connections before the data exchange begins, the TCP continuation, or the termination of the TCP connection. For example, when TCP establishes a connection with three-way handshaking, it will summarise SYN, SNYACK, ACK packets as follows (72095, '0x02', 215.73 sec, 74),(72126, '0x12', 215.78 sec , 70 B), (72127, '0x10', 215.78 sec, 66 B). Then it will follow a stream of packets with TCP flag = 0x10 (ACK) until the connection is disconnected with flag FIN. This will be useful for analyzing any problem with time series or monitoring the discontinuity of such a \pairflow\ as we can see in Section~\ref{sec:ec_profile_features}.

\textit{UDP plane} records all UDP-based packets with protocol name, packet type, timestamp, and packet length. For example, if there are two packets for DNS which are request and response for a specific domain,  they will be summarised as follows: (21160, 'DNS', 'DNS Request', 141.44 sec, 75 B), (21219, 'DNS', 'DNS Response', 141.54, 547 B). 
\textit{ICMP Plane} is follows the same format to the \textit{UDP plane}. However, the type and code are reporting ICMP settings for each packet. The plane can be helpful for any classifier detecting ICMP-based attacks such as scanning activities. 

\subsubsection{FQDN and URL}
As depicted in Figure \ref{fig:pairflow_architecture}, \textit{Domain list} encapsulates all FQDNs related info in a list of tuples. Each tuple holds an FQDN, its A and NS resource records, and the domain age extracted from the WHOIS file. This helps malicious domain detectors, which often rely on FQDN name, relative DNS zone, and WHOIS files. 
\textit{URL} encapsulates each relevant element of URL during a connection in a tuple which includes FQDN, web page filename,the number of parameters, values and fragments, and whether it contains encoded strings or not. 

\subsubsection{HTTP(S)}
\textit{HTTP} encapsulates HTTP-level information for a given connection, in particular, distinct HTTP server names, status codes, content types and User Agents. 
\textit{TLS Protocols} summarises the security settings between a client and server. Cipher suites for both client and server are stored in a list. Cipher suites includes the key exchange/agreement (e.g. RSA, Elliptic-curve Diffie–Hellman (ECDH), Elliptic Curve Digital Signature Algorithm (ECDSA)) , authentication (e.g. RSA), block/stream ciphers (e.g. AES, RC4) with their block cipher mode (e.g. CBC) and message authentication (e.g. MD5, SHA-x). Extension types are also listed for each connection which summarises the cipher suite settings such as extended master secret, session tickets, and Elliptic Curve (EC) point formats. Supported Groups are also stored, known as the EC setting (e.g., secp256r1, secp521r1). 

\subsubsection{Initial Statistical behaviour}
It is important to summarise statistically a few essential fields. We calculate max, min, mean packet TTL, delta packets interarrival time, and duration for a given \pairflow. We also calculate the total (encrypted) bytes and the ratio of sent/received (encrypted) bytes. Max, min, median of cipher suites bytes, and server and client extension bytes are also calculated. We also provide a statistical summary of individual protocol numbers of packets such as raw TCP, raw UDP, ICMP, DNS, HTTP, TLS, and SSL. 

In Table \ref{table:cf_feature}, we summarise all fields produced by \pairflow\ with their datatype. Table \ref{table:earlycrow_feature} presents all features used by \earlycrow\ with their ID to be referred throughout the discussion in Section \ref{sec:EarlyCrowevaluation}. We also mark the features used in the literature and novel ones. 

\begin{table*}[!t]
  \begin{center}
\scriptsize
    \begin{tabular}{p{0.7cm}p{5.5cm}p{0.5cm}p{0.7cm}p{5.5cm}p{0.5cm}} 
      \toprule 
         ID&Field & Type		&	ID& Field & Type		 \\
\toprule
\multicolumn{6}{c}{\textbf{I. Informative Fields}}\\

1& Flow ID  &N&17& HTTP Servers     &LS\\
2-3& Source \& Destination  &S&18&   Status Codes       &LS\\
4& Packet Data Points  &LT&19&  Content Type       &LS\\
5& EPFLAG  &S&20-21& Server and Client Ciphersuit  &LS\\
6-12& EPFLAG raw TCP, raw UDP, ICMP, DNS, HTTP, TLS and SSL  &B&22-23&  Server and Client Extension Type     &LS\\
13& FQDN  &LS&24-25& Client and Server Signature Algorithm and Hash   &LS\\
14& Nameservers Resource Records &LS&26-27& Client and Server Supported Group   &LS\\
15& A Resource Records  &LS&28&  ALPAN Next Protocol &LS\\
16& URL  &LT&29&  EC Point Format &LS\\

\midrule 
\midrule 

\multicolumn{6}{c}{\textbf{II. Statistics Fields}}\\
30& Total Bytes   &N&44-48& TTL Max/Min/Mean/SD     &N\\
31-32& Total Sent/Received Bytes  &N&49-52& Delta Packets  Interarrival time Max/Min/Mean/SD   &N\\
33& Total Encrypted Bytes   &N&53-56&Content Length Total/Max/Min/Median   &N\\
34-35& Total Encrypted Sent/Received Bytes  &N&57-59& Server and Client Cipher Suites Bytes  Max/Min/Median  &N \\
36-42& Number of raw TCP, raw UDP, ICMP, DNS, HTTP, TLS and SSL packets   &N&60-62&Server and Client Extensions Bytes Max/Min/Median&N \\
43& Duration    &N\\

\bottomrule 
    \end{tabular}
  \end{center}
  \raggedright
    \caption{Summary of \pairflow\ data fields (B: Boolean, LS: List of Strings, LT: List of Tuples, N: numerical).}\label{table:cf_feature}
\end{table*}

\subsection{Variants Extraction}

\pairflow\ processing also exports four \emph{variant} JSON files, which can be used by any external classifier, as depicted in Figure~\ref{fig:pairflow_architecture}.
FQDN.json includes all domains and their hostname lists that have been accessed during a given \pairflow. In addition, resource records such as A, NS are also included, and domain age is extracted from the WHOIS file, which appears to be useful for domain detection \cite{alageel2021hawkeye}.
TCP-UDP-ICMP.json is dedicated to those classifiers that use time-series for detection \cite{assadhan2008network,assadhan2009detecting,hu2016baywatch}. All three planes are presented here in addition to related statistical fields such as packet TTL and delta packets interarrival time. 
HTTP.json is employed for those interested in detecting malicious HTTP connections \cite{hu2016baywatch,oprea2018made}. 
Other classifiers may deploy HTTPS.json for detecting encrypted communications without deciphering the traffic \cite{anderson2016identifying}. The following section will discuss how \earlycrow\ used mostly the HTTP variant. A detailed study of the other variants goes beyond the scope of this paper.


\section{\earlycrow}\label{sec:earlycrow_desc}

\earlycrow\ detects malicious HTTP(S) attacks, and in particular APT malware. We now discuss the architecture of \earlycrow, and how features are extracted and updated.

\subsection{Architecture Overview}
\earlycrow\ is composed of four main processes, as depicted in Figure \ref{fig:earlycrow}. 
First, it starts with buffering and dispatching using \pairflow. 
After the \pairflow\ HTTP variant is generated, these flows are preprocessed for profile pivoting to generate three profiles: Host, Destination, and URL. 
Then, two types of feature extraction follow (\pairflow\ and profile features) to form a \contextualsummary\ which is then classified by a random forest. 
When another \pairflow\ is received, it will follow the same workflow. A further step is required when the new \pairflow\ matches one of the previous \contextualsummary\ ID in the repository. The \contextualsummary\ updating process is responsible for updating the matched \contextualsummary\ to maintain the contextualization and reclassify again.
In the following, we describe these four processes briefly.

\subsubsection{Raw Data Buffering and Data Format Transformation}
\earlycrow\ maintains a blacklist of IPs and FQDNs in order to block connections as soon as a malicious entity is detected.
Otherwise, it stores the packets temporarily in a PCAP repository (Figure \ref{fig:earlycrow}, \circled{1}). 
Once the time window has passed, these packets are dispatched to the \pairflow\ module which generates the HTTP variant for the current time window, as explained in Section \ref{sec:cf_description}.

\subsubsection{\pairflow\ Preprocessing}
Incoming \pairflow{}s are preprocessed in two ways. 
First, each \pairflow\ is individually dispatched to Feature Extraction, to extract additional features from the flow perspective (Figure \ref{fig:earlycrow}, \circled{2}). 
Second, a list of \pairflow\ in the same time interval is pivoted to three different profiles.
\emph{Host profile} pivots all \pairflow\ using the source pair (A), \emph{Destination profile} uses the destination pair (B), and  \emph{URL-profile} pivots flows using the FQDN as an identifier. 
These profiles are dispatched to Feature Space Generation.

\subsubsection{Feature Space Generation}
We have two primary dimensions of \pairflow\ to obtain features (Figure \ref{fig:earlycrow}, \circled{3}). 
First, \pairflow\ features help to understand the overall connection (HTTP and DNS) in terms of the statistical behaviour of requests and responses, their temporal analysis, and content exchanged. 
Second, the profile-based features include the three profiles above: Host, Destination, and URL. 
The profile-based features help identify whether a received flow has abnormal behaviour with respect to those profiles. 
For instance, the Host profile may help identify those infected hosts who access two destinations (two different \pairflow{}s) which may linked to the same APT.
Together, they form the \contextualsummary\ (Figure \ref{fig:earlycrow}, \circled{4}), which includes features from different perspectives to help a classifier for accurate classification. 
For example, consider a fallback channel scenario, which is common in APT malware. 
Host $A$ queries a domain and resolves with more than one IP. Hence $A$ may establish multiple unique connections to more than one destination.  
Another case is when host $A$ communicates to remote server $B$, which sends another IP address to communicate in parallel with host $A$, or suspends the first connection and relies on the second as the main channel. 
The \contextualsummary\ catches such behaviour, which would be missed by solutions using exclusively flow-based features.

\subsubsection{\contextualsummary\ Updating Process}
The \contextualsummary\ has an updating process for the next time window for the matching \pairflow\ (Figure \ref{fig:earlycrow}, \circled{5}). First, the updating component inspects the \contextualsummary\ repository to find if a pair has been processed for the incoming \pairflow\ (Figure \ref{fig:earlycrow}, \circled{6}). Then update rules are applied on \contextualsummary\ differently according to the two dimensions i.e. \pairflow\ and profile features (Figure \ref{fig:earlycrow}, \circled{7}).

The rest of this Section discusses in detail the feature space generation and how the \contextualsummary\ is formed.

\subsection{Flow-based Features}\label{sec:earlycrow_pairflow_features} 

\earlycrow\ benefits from using the statistical features produced by \pairflow, which are presented in Section~\ref{sec:cf_description} and Table~\ref{table:cf_feature}. It also extracts higher-level contextual features from the TCP and UDP planes (Figure \ref{fig:pairflow_architecture}, \circled{3}).

\subsubsection{Statistical behaviour}\label{subsubsec:statstical_behaviour}
Identifying the total exchanged bytes can reveal the average for legitimate connections. 
Botnets are typically noisy and have higher total bytes during a flow, while APTs have the lowest total to keep their low profile. %
Another noticeable feature is using raw TCP ratio to detect the \textit{non-application protocols} TTP. 
APTs tend to have the highest, and use HTTP as camouflage while still relying on TCP for many tasks.
Furthermore, identifying the ratio of DNS packets may reveal APTs malicious use of HTTP because it tends to request a domain resolution one time during a connection.
Also, the mean delta packets interarrival time may highlight the low profile of APTs because they are more likely to be significantly longer than legitimate connections.   

From Data Sub-Plane in Figure \ref{fig:pairflow_architecture},
we calculate GET/POST requests and the fraction of status codes started with 1xx, 2xx, 3xx, 4xx, 5xx to identify the most salient behaviour of such a connection. 
It is also essential to analyze the status codes at the packet level to identify whether there are frequent HTTP packets failure due to HTTP server misconfiguration when connecting with APT malware. 
We also notice that measuring the number of resumed connections per flow may highlight the \textit{web protocol} TTP of APT malware.
Using the control sub-plane, we count the termination of TCP connection FIN-ACK (0x11) during a \pairflow\ instead of the sequence of TCP handshaking, i.e., SYN, SYN-ACK, ACK (0x02, 0x12, 0x10) due to the lower computation cost.
However, to exclude a typical HTTP flow (e.g., browsing sessions) for reducing FPR, the number of DNS requests during a given \pairflow, using the UDP plane, may reveal such behaviour.  
\earlycrow\ calculates the number of declarations of content types and their ratios to the others in the data sub-plane.
Examples of considered types: JavaScript, HTML, image, video, application, and text.
APTs and botnets typically have more HTML declarations than legitimate ones, for example due to the use of the \textit{data obfuscation through protocol impersonation} TTP. 
A typical legitimate connection declares the type without the need to redefine it every time unless another content type of a web page is used, such as images or videos. 
Moreover, APTs and botnets rarely use the image content type, which is frequently used in legitimate traffic.

\subsubsection{Time-based behaviour} \label{subsec:time_based}

The challenge of time-based features is to identify APTs connections that operate in low-profile mode. 
First, we consider using a couple of time-based features from the \pairflow\ such as packet TTL, duration of the \pairflow, and delta packets interarrival time, which is the time difference of arrival packets including control, data, UDP, or ICMP packets. 
We can also measure the data packet exchange idle time using the data sub-plane, the difference between subsequent data packets' arrival time. 
We measure the max/min/mean data packet exchange idle time. 
Typical use of HTTP, such as web browsing, has a small gap between delta packet interarrival time and data packet exchange idle time. 
We suspect that APTs may tune data packet exchange idle time to be shorter/faster because they tend to make delta time to be slower to avoid frequent and unnecessary packets.
Once they have to communicate, they send many data packets subsequently, stop for a longer time, and resume later with similar cycles. 
We investigate such APTs approach in Section \ref{sec:EarlyCrowevaluation}.
%

%
We propose additional time-based features that attempt to measure the stealthy behaviour with time-series techniques.
We present features based on the simple moving average (SMA). 
The purpose of the SMA is to average the data points over a time window of size $t$ decided in advance, so that an analyst can identify when a data point is above or below such average. 
$SMA_k$ can be described as follows:
\begin{equation}
    SMA_k= \frac{1}{k} \sum_{i=n-k+1}^n{p_i} 
\end{equation}
where $p$ is the packet length, $k$ is the number of previous data points in a time window, and $n$ is the current data point. 
Since packets arrive asynchronously, in order to calculate an S$MA$ we need to introduce a sampling rate such that packets arrived within two sampling events are combined together in a single point.
For example, if the time window is one minute and we sample points every second, then $k=60$ and if we receive two packets of length, respectively 128 and 32, between seconds 5 and 6, then $p_6=160$. 
After calculating the SMA, we can extract the number of outliers and their ratio and magnitude. 
Outliers are those points two times above the corresponding $SMA_k$. 
Therefore, we can capture the stealthy behaviour of APTs, which has fewer outliers than legitimate and botnet traffic.
However, it is also essential to find the number of packets below and above average.
These features can capture the APTs that touch or slightly exceed the $SMA$, reflecting cautious operation.

\begin{figure*}[!t]
\centering
\includegraphics[width=18cm,height=8cm]{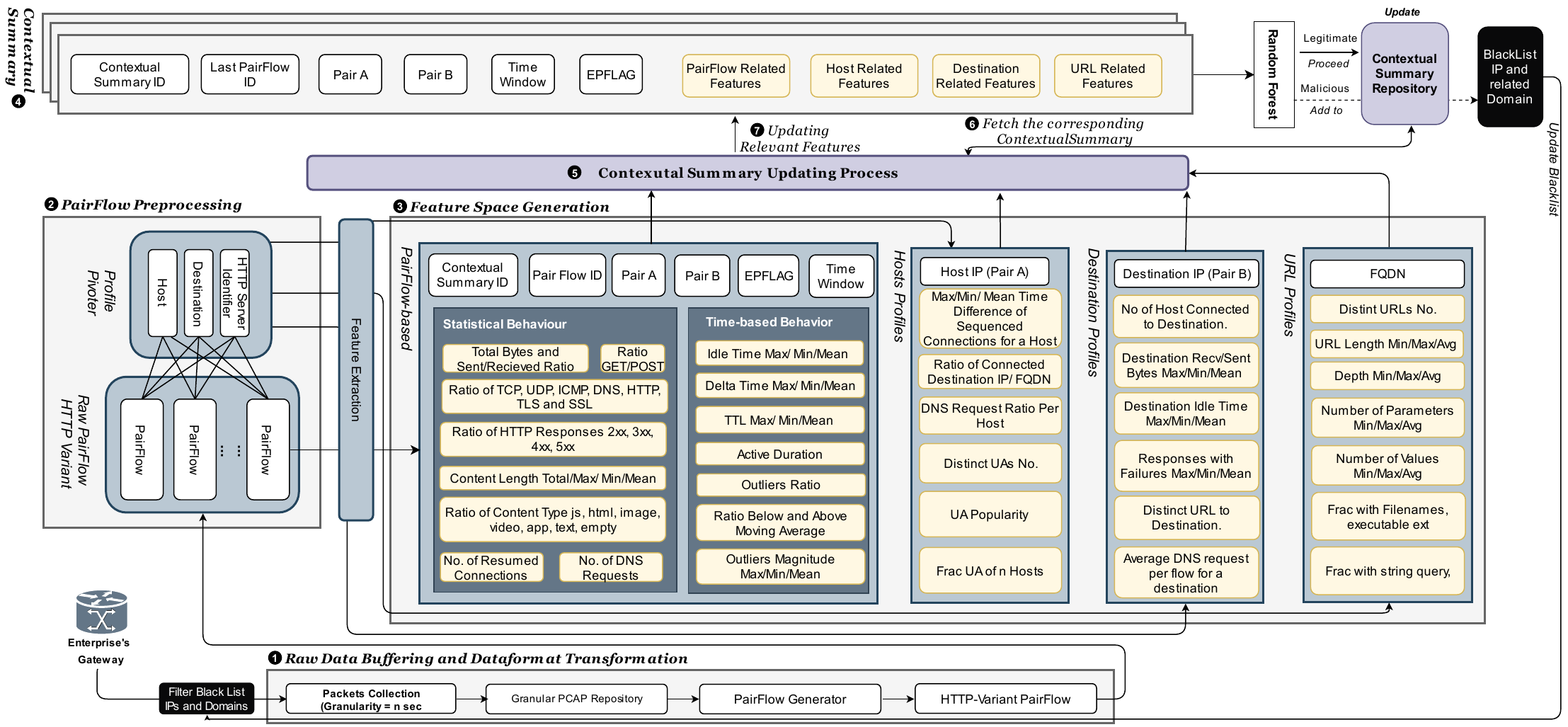}
\caption{Overview of the \earlycrow\ architecture.}\label{fig:earlycrow}
\end{figure*}

\subsection{Profile Features} \label{sec:ec_profile_features}

Profiles features are generated based on all \pairflow{}s with longer time windows, for example lasting several weeks, or even months. 
\earlycrow\ queries the related info using a host IP, destination IP, and FQDN for Host profile, destination, and URL profile, respectively. %
The purpose of the host profile is to identify whether that host has a sign of infection such as discrepant information or fallback channel for example.
The destination profile may reflect those destinations that an enterprise can access and avoid some false positives.
The URL profile helps identify the typical use of a given FQDN. 
FQDNs commonly accessed without parameters or values, especially with GET as method type, could signal use of the Dynamic DNS technique to point to frequently changed IP addresses known to be used for APTs \cite{alageel2021hawkeye}.
Long URLs with many parameters and values for a given FQDN could instead be legitimate URLs because many hosts request different URLs, indicating rich website access. 
The URL profile helps pinpoint the malicious use of HTTP protocol from their past behaviour.
Yet, APT cannot be easily detected based on such single features, so these will only contribute in part to the final classification.

\subsubsection{Host Profile} 

The purpose of the host profile is to investigate the effect of infection on a machine behaviour over $\hat{t}$ time which should be longer than the selected granularity $t$ time for \pairflow.
Benign hosts should have specific characteristics in terms of resumed connections, DNS requests per flow, time difference of sequence connections, type of UA used. 
When a host is infected with APT malware, its characteristics may move to another point further from the benign host centroid. 
For instance, it is suspicious for a host to initiate a connection by IP only, which is highly linked to a \textit{fallback channel}.

\begin{table*}[t!]
  \begin{center}
\scriptsize
    \begin{tabular}{p{0.7cm}p{5.7cm}p{1.2cm}p{0.7cm}p{4.9cm}p{1.2cm}} 
      \toprule 
         ID&Feature 	&New? &	ID& Feature 	&New? \\

       \midrule 
       \midrule 
\multicolumn{6}{c}{\textbf{I. \pairflow\ Based Features}}\\
1 &Total  Bytes  &\cite{bartos2016optimized,bilge2012disclosure,tegeler2012botfinder}&28-31&  No and ratio below and above AVG   &\checkmark\\
2 &Sent/Received Ratio   &\cite{bartos2016optimized,bilge2012disclosure,tegeler2012botfinder, oprea2018made}&32-33&   No and ratio outliers  &\checkmark\\
3-9 &Ratio of raw TCP, raw UDP, ICMP, DNS, HTTP, TLS and SSL packets   &\checkmark&34-37&  outliers Magnitude Max/Min/Mean/SD    &\checkmark\\
10-13 &Ratio of HTTP Response packets with 2xx, 3xx, 4xx, 5xx   &\cite{oprea2018made}&38-40&Data packet exchange idle time Max/Min/Mean    &\checkmark\\
14-15 &Ratio of frequent GET and Post   &\cite{perdisci2010behavioral,oprea2018made}&41& Active duration  &\checkmark\\

16-19 & Content Length Total/Max/Min/Median   &\checkmark&42-45 &Packet TTL Max/ Min/ Mean/ SD   &\checkmark\\
20-26 & Ratio of Content Type Javascript, HTML, Image, Video, App, Text, Empty     &\cite{oprea2018made}&46-49 & Delta packets interarrival time Max/ Min/ Mean/ SD   &Similar to \cite{bartos2016optimized}\\

27 & No of resumed connections   &\checkmark&50 &No of DNS request   &\checkmark\\

       \midrule 
\multicolumn{6}{c}{\textbf{II. Host Profile Features}}\\
51-53 & Max/Min/Mean Time Difference of Sequenced Connections  &\checkmark&59 & Distinct UAs per host      &\checkmark\\
54 & Ratio of Connected Destination IP only to FQDN  &\checkmark&60 & nAvg UA Popularity    &\cite{oprea2015detection}\\
55-57 &  Max, Min, Mean of Resumed Connections per flow for a host      &\checkmark&61-62 & Frac UA 1, 5   &\cite{oprea2015detection}\\
58 &   No of DNS request per flow for a host    &\checkmark&63 &  Ratio UAs   &\cite{oprea2015detection}\\

       \midrule 
\multicolumn{6}{c}{\textbf{II. Destination Profile Features}}\\
64 & No. of hosts connected to destination  &\cite{oprea2015detection}&72 &   No. of Distinct URLs associated to destination&\checkmark\\
65-67 & Destination Received /Sent Max/Min/Average  &\checkmark&73-75 & Destination Max/Min/Mean  Packets Failure  &\checkmark\\
68-70 & Destination Data Packet Exchange Idle Time Max/Min/Mean  &\checkmark&76-81 & Max/Min/Mean No and ratio of DNS request per flow for a destination   &\checkmark\\
71 & No of resumed connections per flow for a destination     &\checkmark&\\

\midrule 

\multicolumn{6}{c}{\textbf{II. URL Profile Features}}\\

82 & Frac URLs filename  &\cite{oprea2018made}&94-96 & URLs Values Max/Min/Mean   &\cite{bartos2016optimized,perdisci2010behavioral,oprea2018made}\\
83 & Frac URLs filename exe  &\checkmark& 97-99 &URLs Fragments Max/Min/Mean   &\cite{oprea2018made}\\
84 &Number of distinct extensions &\cite{oprea2018made}  &100 &Frac query   &\cite{oprea2018made}\\
85-87 & URLs Length Max/Min/Mean  &\cite{bartos2016optimized,perdisci2010behavioral,hu2016baywatch,oprea2018made} &101 &Num hasString   &\checkmark\\
88-90 & URLs Depth Max/Min/Mean   &\cite{bartos2016optimized,perdisci2010behavioral,oprea2018made}&102 &Num of URLs and Distinct ones   &\cite{invernizzi2014nazca}\\
91-93 & URLs Parameters Max/Min/Mean   &\cite{bartos2016optimized,perdisci2010behavioral,hu2016baywatch,oprea2018made}\\

\bottomrule 
    \end{tabular}
  \end{center}
  \raggedright
    \caption{\earlycrow\ features. Note that features reused from the literature are computed from \pairflow\ data rather than from other data formats.}\label{table:earlycrow_feature}
\end{table*}


\earlycrow\ investigates the number of resumed connections per flow for each host. 
Legitimate HTTP usage typically increases the number of resumed connections because each time a web page is downloaded, the TCP connection is terminated with FIN. 
Then, when a host clicks on another link, even for the same website, a new TCP three-way handshake is initiated and completed, followed by a connection termination once the page is completely retrieved. 
The scenario remains similar for a web caching but with a different destination when a host contacts a proxy server or content delivery network (CDN). 
Hosts infected with APTs have lower resumed connections per flow than benign hosts.
APTs tend to terminate less because HTTP(S) usage is for malicious use, and to stay stealthy, they attempt to avoid frequent connection termination.
However, botnets have higher connections terminations than the legitimate and APTs due to their noisy nature. 
Similarly, we extracted the DNS request per flow to identify a host with lower DNS requests than expected, which is also a sign of APTs using dynamic resolution, DGA and data obfuscation through protocol impersonation TTPs. 
Hosts infected with botnets are likely to use excessive DNS requests even more than the legitimate ones because of DGA techniques \cite{antonakakis2012throw} or \cc\ communication over DNS \cite{dietrich2011botnets}. 

%
%
We also measure the Mean Time Difference of Sequenced Connections (MTDSC), which can help to identify \textit{fallback channel}. 
MTDSC can be calculated as follows:
\begin{equation}
    MTDSC = \frac{1}{n} \sum_{i=0}^n{t_{i+1}-t_{i}} 
\end{equation}
where $n$ is the number of new connections and $t$ are their timestamp.
The input timestamp should be the first packet sent or received from Control, UDP, or ICMP planes for any \pairflow, where the source is the same host.
Hosts infected with APTs may have a higher mean for MTDSC than those infected by botnets and benign hosts.
They initiate new connections after a long time for their \textit{fallback channel}, which affects the MTDSC even if a host has many normal browsing connections.
%

We also compute a ratio of connected destinations using IP only to those with FQDN.
The feature can capture the APTs behaviour of using DNS requests to locate the IP address of \cc; once the first channel is established, APT malware sends another IP as a fallback channel and starts another three-way handshake.
Botnets are less likely to use the same approach, while it is infrequent for legitimate HTTP to be accessed by IP only. 
%
%
As pointed out in Section \ref{sec:EarlyCrow_threat_model}, any client that uses HTTP will have an optional UA in a request packet, and it could be a (non-)browser, malicious string, or just an empty.
Similar to \cite{oprea2018made}, we extract several features for UA, including the distinct number of UAs and their popularity among an enterprise. 
%

\subsubsection{Destination Profile} 

The destination profile analyzes the servers contacted by internal hosts to find the characteristics of the provided services.
We are interested to find out if it is normal for a destination to have fewer/higher DNS requests, short/long data packet exchange idle time, high/low packet failure, sending/receiving dominant, and high/low resumed connections.  
%
%
For instance, we measure the number of DNS requests per flow for a destination to investigate if such destination is using \textit{dynamic resolution, DGA or data obfuscation through protocol impersonation} TTPs. 
An APT destination usually has fewer DNS requests than usual. 
Once the domain is resolved and TCP establishment has been completed, it is rare to request more DNS packets. 
The legitimate use of HTTP(S) is to query the DNS packet every time they visit each page. 
Therefore, the number of DNS requests is directly proportional to HTTP packets.
%
%
%

It is also essential to measure the destination data packet exchange idle time to identify legitimate web servers with a reasonable time to be idle for browsing.
The typical APTs destinations have lower data packet exchange idle time than both legitimate and botnets due to the need for continuous or beaconing communication with lower outliers over a long time.
Again, the data packet exchange idle time here focuses only on the meantime of zero data exchange packets from a destination point of view without considering the control ones. 
%

%
As pointed out in Section \ref{sec:EarlyCrow_threat_model}, some APTs use \textit{protocol impersonation} such as HTTP as camouflage to communicate with \cc. 
Thus, identifying the packet failure for each destination is essential to find if the failure comes from the destination itself here defined in this section or from the \pairflow\ in Section \ref{subsubsec:statstical_behaviour}. 
The objective is to find if a destination mimics web browsing activities while mainly communicating with the victims with raw TCP as \textit{non-application protocol}.
Therefore, many APTs destinations have server misconfiguration due to that fake need, causing higher packet failure (response with 4xx or 5xx) than legitimate.
%
%
We also measure the destination received/sent bytes ratio. 
APT malware tends to send instructions and payloads to the infected machine more than data exfiltration at the beginning of the APTs cycle. 
We measure each destination's receive/sent bytes ratio to identify the APT, and we find that the infected machine has a higher destination ratio than the legitimate ones. 
On the contrary, botnets tend to exfiltrate data more than sending payloads and instructions. 

\begin{table*}[!t]
  \begin{center}
\scriptsize
    \begin{tabular}{p{1.7cm}p{1.7cm}p{3cm}p{9cm}} 
      \toprule 
         \textbf{Label}&\textbf{\#Packets }& \textbf{Set}&\textbf{Malware Families 	}
         \\
\toprule
\multirow{4}{*}{\textbf{Malicious}}& \multirow{4}{*}{567,090} &Training& Bitsadmin (0.09\%), Carbank (0.05\%), Conficker (27.56\%),  Mivast\&Sakula (0.93\%), NanoCore (0.13\%), njRAT (28.45\%),  PlugX (0.11\%) , Remcos (0.87\%),  Sogou (3.65\%), Virut (9.59\%), Zebrocy (0.98\%),
 \\

&&Testing (Unseen) & Ammyy (1.01\%),  ChChes (0.13\%), CobaltStrike (0.39\%), Dridex (0.23\%), Emotet (0.02\%),  Empire (1.70\%),  FlawedAmmy (0.24\%),  ImminentMonitor (11.27\%), MagicHound (0.40\%), OnionDuke (0.14\%), PoisonIvy (0.25\%),   Ramnit (0.21\%), StrongPity (11.38\%), Zeus (0.04\%)

\\

\textbf{Legitimate}& 766,641 & - & Training: 70\%, Testing: 30\%.\\

\bottomrule 
    \end{tabular}
  \end{center}
  \raggedright
    \caption{Dataset characteristics used for measurements (training set) and for unseen malware evaluation (testing set).}\label{tab:EarlyCrowdataset}
\end{table*}
%
Another important aspect for each destination is calculating the number of resumed connections.
Browsing behaviour has frequently more resumed connections than the APT ones.
Each time a user moves to another page, three-way handshaking is established, incrementing the destination of the resumed connection by one.
On the APT, once the connection is completed, it is rare to disconnect the communication and resume it later to avoid noisy TCP handshaking. 
%
%
Another important feature is to observe the number of hosts connected to each destination.
Popular web servers and botnets destinations are routinely contacted by a considerable number of hosts.
In contrast, APTs typically infect as few as possible hosts, hence receiving few connections to their destinations.

\subsubsection{URL Profile} 

We present URL-based features which are separate from those in the destination profile, as many FQDN-based URLs share the same IP or vice versa, host several different IPs. The URL profile summarises the standard behaviour of URI resources and the traffic statistics for each FQDN or IP-based URL.  
We count here how many URLs are reached during a connection and how many are distinct.
A malicious  \cc\ server typically has fewer than a legitimate one.
However, in case of evasive use of \textit{web protocol}, APTs may have more distinct than botnets. 
We also check if a URL has a query string, filename, and whether it has an executable extension, then calculate the fraction of the number of each field compared to the distinct number of URLs. 
A legitimate URL is likely to possess a filename with a variety of extensions. 
Other statistical features, i.e., Min/Max/Mean, are also calculated on URL length, depth, number of parameters, values, and fragments. 
The legitimate URL Profile is typically the longest because it has more depth, parameters, and values. 
APTs come next, followed by botnets.

\subsection{\contextualsummary} \label{sec:contextual_summary}

When all features are extracted from \pairflow, and profile-based features are prepared, the \contextualsummary\ module collects these features in one bundle to be dispatched to the classifier.
When a new \pairflow\ is received, \earlycrow\ checks the \contextualsummary\ repository to identify if the pair had been already processed in the past.
If so, the \pairflow\ will be processed as described in the previous sections.
Then, it will be dispatched to the updating process module to combine the new flow with the previous ones as described in the next section.
The purpose is to track the same connection over time to catch malicious behaviour.
For example, if a malicious actor bypasses \earlycrow\ for the first flow, it will be tracked over time until it gets blocked.
Indicators associated to positive detections may stay in the \contextualsummary\ repository and the blacklists for training the classifier. 
In Table \ref{table:earlycrow_feature}, we summarise all features included in \contextualsummary. 


\subsection{\contextualsummary\ Updating Process}

While \pairflow{}s are stored in a repository, the \contextualsummary\ gets updated over time, using different rules for Host, Destination, and URL Profiles. 
If an incoming \pairflow\ has no associated \contextualsummary, a new one is created. 
Otherwise the new \pairflow\ is considered for feature extraction, causing an update of the corresponding features of the associated \contextualsummary.
The time window is expanded with the new \pairflow\ to describe the overall time window covered by the \contextualsummary. 
However, updating profile-based features could cause higher time complexity because these profiles are to be updated for every different \contextualsummary. 
Therefore, new profile-based features are recalculated every $\hat{t}$ time, such that $\hat{t} > t$, where $t$ is the selected granularity for \earlycrow. 
For instance, we can configure $\hat{t}$ at 15 minutes in our experimental settings, which is higher than $t$ by 50\% if $t$ at 10 minutes.

\earlycrow\ considers different methods to update features according to their data type. 
Numerical features are updated by using a weighted average.
As shown in Figure \ref{fig:earlycrow}, each \contextualsummary\ stores the last \pairflow\ ID as a counter of previous ones to be used for the weighted average formula.
EPFLAG-based features, Boolean data type, are updated with OR operation with an incoming one to summarise the overall protocol used during \contextualsummary.
For instance, APTs often have the DNS packets at the first \pairflow, but not the subsequent one, as we discuss in Section \ref{sec:EarlyCrow_threat_model}.
Therefore, updating the \contextualsummary\ does not reset EPFLAG for the DNS.
For Host-Profile features, strings of UA are stored to accurately extract other related UA, such as the number of distinct UAs which cannot be updated without having access to their strings.

\begin{table*}[h]
  \begin{center}
\scriptsize
    \begin{tabular}{p{2.7cm}p{13cm}} 
      \toprule 
         Malware & Used in 	
         \\
\toprule
Ammyy & FIN6 and TA505 \\

ChChes&APT10 \\

Cobalt Strike & APT10, APT19, APT29, APT32, APT37, APT41, DarkHydrus, Chimera, Cobalt Group, CopyKittens, FIN6, FIN7, Indrik Spider, Leviathan, Mustang Panda and Wizard Spider  \\

Empire &  APT19, APT33, APT41, CopyKittens, FIN10, Frankenstein, Indrik Spider, Leviathan, MuddyWater, Silence, Turla,  Wizard Spider and WIRTE,  \\

Imminent Monitor & APT-C-36\\

Magic Hound & APT35\\

MiniDuke &APT29\\

Mivast \& Sakula &Deep Panad\\

NanoCore &  APT33, Gorgon, Group5 and SilverTerrier\\

njRAT &  APT41, Gorgon, Group5 and Transparent Tribe\\

OnionDuke &  APT29\\

PlugX &  APT3, APT10, APT27, APT41, DragonOK, GALLIUM, Higaisa, Mustang Panda and TA459 \\

PoisonIvy &  admin@338, APT1, APT10, DragonOK, Elderwood, GALLIUM, IndigoZebra, Moafee, Molerats, Mustang Panda,  PittyTiger,  Soft Cell and Tropic Trooper\\

Remcos &  Gorgon\\

StrongPity & PROMETHIUM\\

Zebrocy &  APT28\\

\bottomrule 
    \end{tabular}
  \end{center}
  \raggedright
    \caption{A list of potential campaigns using APT malware presented in APTraces and MCFP.}\label{tab:EarlyCrowCampaigns}
\end{table*}


\section{APT Malware Dataset}\label{sec:datasets}

APT malware attacks a few targets in discontinuous time-frames spanning months or years, unlike other malware and common attacks. Therefore, the chance of finding a real network infected with various APT campaigns is unrealistic.
We resort to raw PCAP captures from two different honeypot networks, each of which includes legitimate, APTs and botnets \cc\ connections (Table \ref{tab:EarlyCrowdataset}). 
After \pairflow\ compiles the PCAPs and generates HTTP variant files, we build three combined datasets: APTs vs. Legitimate, botnets vs. Legitimate, and Malicious (APTs or botnets) vs. Legitimate. 
We then causally analyze these datasets and select appropriate mitigation techniques for bias and spurious correlations. 

\subsection{Captures} \label{page:earlycrow_dataset}
\subsubsection*{\textbf{APTraces}:} 
We run different active APT malware using Any.Run\footnote{https://any.run} sandbox machines to generate PCAP files. 
These malware families are known to be used by 48 APT campaigns\footnote{Campaigns use each APT malware can be found at https://attack.mitre.org/groups/} (Table~\ref{tab:EarlyCrowCampaigns}). However, they are often temporarily inactive. Due to this, we run them during multiple time windows (October - November 2019, April 2020 - January 2021) until each campaign's activities are resumed, and their command and control are activated.
Malware families included in this dataset are RATs (\textit{njRAT, Imminent Monitor, CrossRAT, Mivast \& Sakula, NanoCore, PlugX, PoisonIvy}) and trojans (\textit{Empire, OnionDuke, MiniDuke, Remcos, StrongPity, Zebrocy}).

We also consider legitimate connections from the same sandbox to avoid data bias based on the victim machine, configuration settings, or temporal bias \cite{arp2022and} against legitimate.

\subsubsection*{\textbf{Malware Capture Facility Project (MCFP)}:}
The MCFP\footnote{https://www.stratosphereips.org/datasets-overview} is a repository of 349 public captures of different malware families provided by Stratosphere Research Laboratory. We select malware used in APTs such as (\textit{Magic Hound and Cobalt}), admin tools (\textit{Ammyy}), and RATs (\textit{njRAT}).
We also add botnets captures that use HTTP(S) for \cc\ communication  (\textit{Conficker, Dridex, Emotet, Ramnit, Sogou, Virut, Zeus}) and normal traffic (CTU-Normal-12, 20-22).
%
%

\subsection{Causal Analysis and Control Measures} \label{appendix:causal}
Causal analysis is a recommended approach to make sure of data quality in machine learning, especially for predictive tasks. In this section, we diagnose our dataset based on the causal analysis and provide the control measures. We follow D. Castro et al. \cite{castro2020causality} work for the definition of causal analysis and methods to ensure data quality for our dataset.

\subsubsection{Causal Analysis}
\begin{figure*}[!t]
\centering
\includegraphics[width=13cm,height=2.5cm]{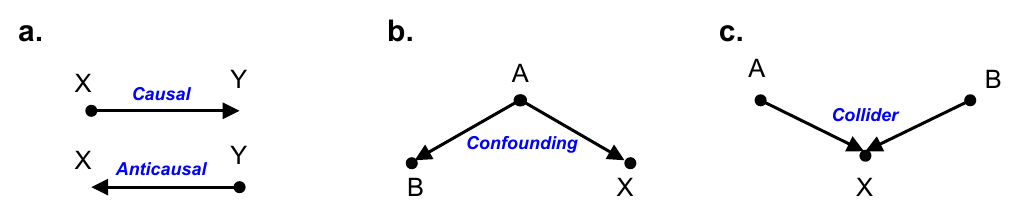}
\caption{Causal reasoning relationships \cite{castro2020causality}.}\label{fig:causal_def}
\end{figure*}

\begin{figure}[!t]
\centering
\includegraphics[width=9cm,height=9.3cm]{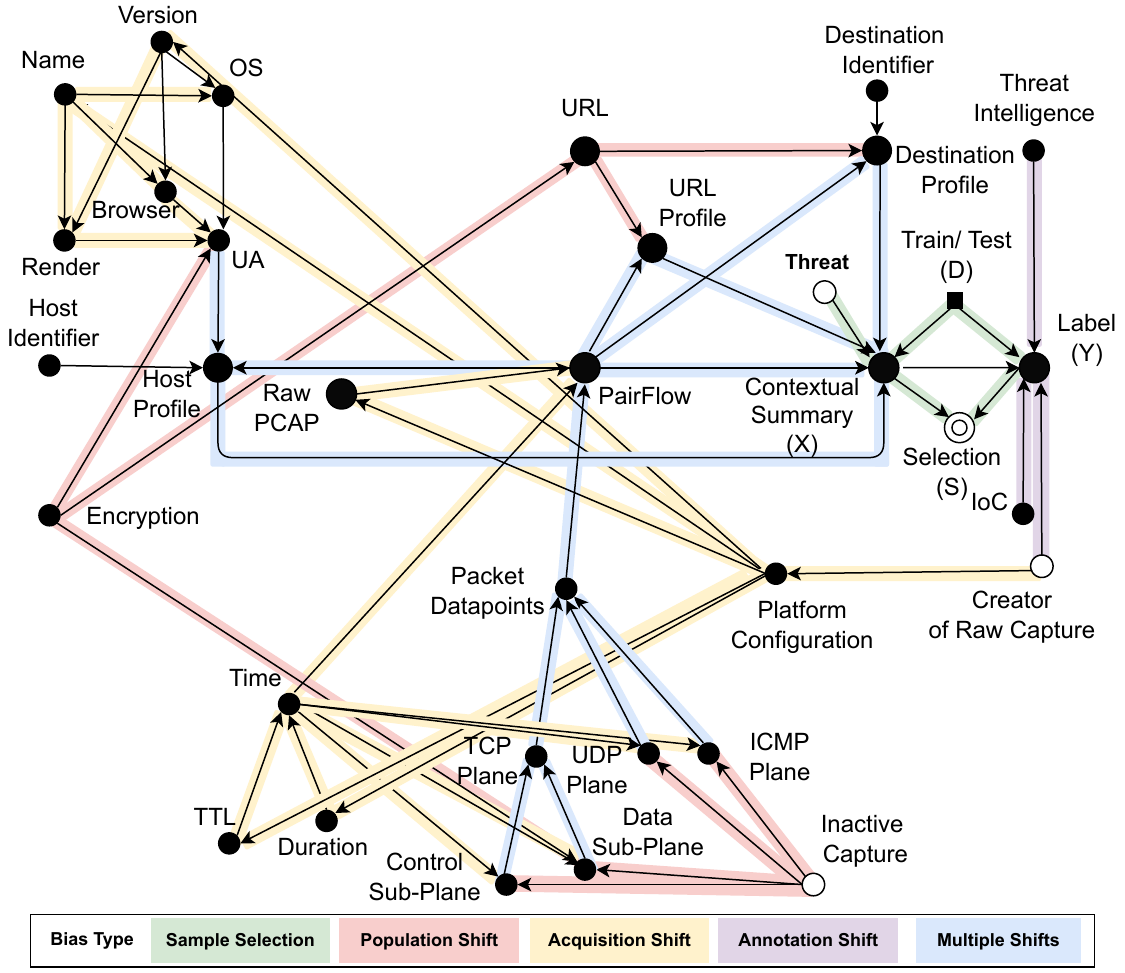}
\caption{Causal model diagram. Empty circles denote hidden variables, whereas solid circles refer to explicit ones.}\label{fig:causal}
\end{figure}

A causal diagram is a directed acyclic graph (DAG) that represents the cause-effect relationship between variables considered in a predictive task, and helps elicit or mitigate bias from experiments. 
A link between a pair of edges can be causal or anticausal. A causal relationship is described by $P(Y|X)$ when $X \rightarrow Y$, where $X$ is a sample and $Y$ is the label. Literally, '$X$ is a direct cause of $Y$, which means modifying the value of $X$ will change the likelihood of $Y$ (Figure \ref{fig:causal_def}.a). Causal links refer to predicting the effect ($Y$) from the cause ($X$). In contrast, anticausal, $P(Y|X)$ when $Y \rightarrow X$, meaning anticausal links predict cause from the effect such as predict $X$ (sample) from $Y$ (label) as depicted in Figure \ref{fig:causal_def}.a. Clearly, anticausal is out of our scope because it does not influence the data-generating process of our dataset, which will use $X$ (\contextualsummary)  only on predicting $Y$ (APTs, botnets and legitimate). 
Therefore, we build a causal model focusing on causal links as depicted in Figure \ref{fig:causal}. We attempt to balance between accuracy and clarity on the level of abstraction. We aim to identify the bias, including confounder and collider, in our dataset and control them. We will explain Figure \ref{fig:causal} after introducing some related concepts.

Confounder may be an unadjusted common cause, while collider involves conditioning a common effect such as selection bias. 
Let us consider that $A$ (say, \pairflow\ ) is a common cause of $B$ and $X$, a Destination Profile and a \contextualsummary, respectively (Figure \ref{fig:causal_def}.b). $A$ is called a confounding factor, and it creates a spurious correlation between $B$ and $X$, resulting in  $B \not\!\perp\!\!\!\perp X$  (literally, "$B$ is dependent on $X$"). However, they become independent when $A$ is controlled: $B \independent X | A$. 
Consider the instance when $X$ is a common effect of both $A$ and $B$ (Figure \ref{fig:causal_def}.c). We call $X$ as a collider in this case. Conditioning on $X$ adds a connection between $A$ and $B$, as they can now explain away the influence of each other on the observed result, X (i.e., $A \independent B | X$). 

Due to the various capture environments for our considered dataset, it is vital to control collider and confounder caused by data shift. The data shift in causal direction can be population, annotation, acquisition shifts, or sample selection bias. 
Let us assume $Z$ is a vector of the PCAP raw features. \textit{Population Shift} may occur when  $P_{train}(Z) \neq P_{test}(Z)$. We ensure to mitigate such bias by focusing on samples that belong to the same protocol - in our case HTTP(S) - have to be used in each sample. 
\textit{Annotation Shift} is given by $P_{train}(Y|X) \neq P_{test}(Y|X) $ where samples are labelled based on different metadata. For example, APTrace may include a malicious flow that is also presented in MCFP with a different label source. In this case, control measures can be taken, such as relabelling all flows by referring to the same security intelligent platform source. 

\textit{Acquisition Shift} can occur due to different platform configurations, which may affect several features, including packet TTL, duration, and UA-based features. Data harmonization, such as extracting domain-invariant representations, is the foremost approach to mitigate such bias to validate $\vec{F} \independent P_f$, where $\vec{F}$ is a feature vector, and $P_f$ is the platform.
\textit{Sample Selection Bias} is when $P_{train}(X,Y) \not\equiv P_{test}(X,Y) $, meaning the training set does not represent the target population. We can avoid it with a random sample selection with stratified training and testing split for each class.

In Figure \ref{fig:causal}, we point out all possible cause and effect links, where some are shaded with specific colours reflecting the type of potential bias as annotated in the legend box. 
$Y$ (Label) is a common effect of three causes, threat intelligence, IoC, and creator of raw capture. $Y$ is a collider, and the potential bias is the \textit{annotation shift} because we may have conflict among labels based on different sources. 
The platform configuration is a common cause of UA, Time, and Raw PCAP variables. In this case, we call the platform configuration variable a confounder that produces an association between all three variables. The type of shift between these variables is \textit{acquisition shift} due to the different platforms that we collect our captures from. 
The encryption variable is also a confounder. It is a common cause of UA, URL, control, and data sub-planes. For instance, UA and URL are hidden inside the payload when encryption is applied. The details of the data sub-plane are not available without decryption, leaving the data plane with SSL/TLS packets. The potential bias is the \textit{population shift}. Another population shift is caused by the inactive capture variable, which is also a confounder of control, data, UDP, and ICMP planes. When a capture is collected, but there is no sign of attacks or communication with \cc\, it affects the behaviour of these planes to act as legitimate or relatively low number of malicious packets exchange. When two or above shifts are possible, we call such a link as \textit{multiple shifts}. However, mitigating the individual bias will reduce the multiple shifts.

\begin{table*}[t!]
\begin{center}
\scriptsize
    \begin{tabular}{p{4cm}p{0.8cm}p{0.8cm}p{0.8cm}p{0.2cm}p{0.8cm}p{0.8cm}p{0.8cm}p{0.2cm}p{0.8cm}p{0.8cm}} 
      \toprule 
      & \multicolumn{3}{c}{\textit{Macro}}&& \multicolumn{3}{c}{\textit{Weighted}}\\
    \cmidrule{2-4} \cmidrule{6-8}

         Classifier Name	&  \mpscore & \mrscore  & \mfscore &	& \wpscore & \wrscore  & \wfscore& &$FPR$ & $A$\\
    \midrule 
    \midrule 
          
          \multicolumn{10}{c}{\textbf{I. Dataset: APTs vs. Legitimate}}\\
         \earlycrow\	&\textbf{94.20}&	\textbf{93.69}&	\textbf{93.89}&&	\textbf{97.55}&	\textbf{98.41}&	\textbf{98.21}&&	\textbf{0.40}&	\textbf{99.17}  \\
         Baseline	&92.40&	89.09&	90.45&&	95.52&	98.35&	96.85&&	0.49&	98.75 \\
         \earlycrow-HTTPS	& 92.85&	93.18&	92.79&&	97.98&	98.34&	96.53&&	0.51&	99.03	  \\
         Baseline-HTTPS	& 82.90&	68.96&	73.18&&	91.81&	95.31&	93.04&&	0.72&	97.19  \\

          \multicolumn{10}{c}{\textbf{II. Dataset: Botnets vs. Legitimate }}\\   
         \earlycrow\	& \textbf{96.49}&	\textbf{95.40}&	\textbf{95.90}&&	\textbf{99.23}&	\textbf{99.50}&	\textbf{99.36}&&	0.48&	\textbf{98.92}	  \\
         Baseline	& 94.64&	86.92&	90.24&&	98.23&	98.91&	98.53&&	0.57&	97.61  \\
         \earlycrow-HTTPS	&96.79&	95.02&	95.84&&	98.55&	99.47&	99.00&&	\textbf{0.42}&	98.92	  \\
         Baseline-HTTPS	& 90.73&	80.76&	84.79&&	96.10&	98.66&	97.28&&	0.96&	96.39  \\

        \multicolumn{10}{c}{\textbf{III. Malicious vs. Legitimate}}\\   
         \earlycrow\	& \textbf{95.41}&	\textbf{94.79}&	\textbf{95.06}&&	\textbf{97.59}&	\textbf{98.18}&	\textbf{97.87}&&	\textbf{0.86}&	\textbf{98.29}	  \\
         Baseline	& 93.76&	88.20&	90.68&&	94.52&	97.38&	95.84&&	0.93&	96.97  \\
         \earlycrow-HTTPS	& 95.07&	94.76&	94.89&&	97.09&	98.17&	97.62&&	0.93&	98.23	  \\
         Baseline-HTTPS	& 91.21&	78.66&	83.47&&	90.60&	95.53&	92.57&&	0.95&	95.13	  \\

  \bottomrule\\\\[-10pt] 
    \end{tabular}
     \end{center}

   \raggedright
   \caption{Known malware classification performance.}\label{tab:results_known_http_malware}
\end{table*}

\subsubsection{Causal Control Measures}\label{subsec:control}

%
We take several control measures to improve our datasets motivated by the causal model reported in Figure \ref{fig:causal}. 
\textit{Population shift:} 
We exclude any flows that do not use HTTP from the inputs to the \pairflow-HTTP variant.
%
%
%
We limit all captures to the same duration for the duration-related features to mitigate the bias of packet statistics and the duration of a flow itself. 
For instance, it is biased to compare a flow captured for 15 minutes to the other during 2 hours, which may influence the statistics for the benefit of a longer capture. 
We also consider recently recommended practices to troubleshoot datasets for IDS \cite{engelen2021troubleshooting}. 
Therefore, we removed samples without data transfer or failed TCP connections to \cc\ servers.
\textit{Annotation Shift:}
The sandbox already generates an IoC file in APTrace.
However, we label those missed via sandbox by manually checking every IP and domain using other threat intelligence platforms and APTs industry reports following the recommendation by \cite{arp2022and}.
For MCFP, we label the flows based on the original project and manually label them using historical data retrieved from security intelligence platforms, including \textit{IBM X-Force Exchange}\footnote{https://exchange.xforce.ibmcloud.com} and \textit{AlienVault Open Threat Exchange}\footnote{https://otx.alienvault.com}.
%
%
Finally, we need to mitigate the bias for the Host profile in \earlycrow\ through various captures.
Instead of considering source IP as a host identifier, we used a tuple of $(multiple \; labels,source)$ to avoid mixing data with the same internal IP but in a different network or capture.
The multiple label field refers to the filename of a PCAP capture. 

\textit{Sample selection bias:}
To mitigate the bias caused by APTs only as a malicious class, we added botnets \cc\ captures to measure if our system can detect both activities.
Then we stratified the train and test process to ensure $P_{train}(X,Y) \equiv P_{test}(X,Y)$. 
In Section \ref{sec:EarlyCrowevaluation}, we report the weighted and macro average F score due to the binary labels' imbalanced samples and shed light on malicious flow detection. 

\textit{Acquisition Shift:}
The differences in platform configuration cause bias.
We consider only features with domain-invariant representations to ensure  $\vec{F} \independent P_f$.
Therefore, we neglect features on names and versions of OS, Browser, and renderer.
We keep the distinct number of UA per host and its popularity-based features.
Packet TTL is another domain variant, which is dependent on the platform.
We omit features based on TTL, although we kept it available in \pairflow\ for the research community but not for our assessment in the next section.

\section{Results}
\label{sec:EarlyCrowevaluation}

This section evaluates \earlycrow\ performance on the three datasets described in Section \ref{sec:datasets}, each with a split of 70\% for training and 30\% for testing.
The same experiments are performed on a baseline, inspired by \made~\cite{oprea2018made}, which is a NIDS detecting \cc\ used by botnets, ransomware, and APTs. Since we focus on detecting Case I and II in Section~\ref{sec:EarlyCrow_threat_model}, we omit inapplicable features, such as those related to domains, that would be missing for Case II. 

Classifiers are evaluated in two modes. 
First, HTTP-Mode, which assumes the administrator connects the NIDS to a web proxy to decrypt HTTPS and accesses features such as UA, HTTP response codes, content type, and URL.
Second, HTTPS-Mode, where the administrator places the NIDS at the network edge, without deciphering HTTPS.
Because of imbalanced classes of APT (3.9\%) and botnet (8.3\%) compared to legitimate, we focus on macro average F1-score (\mfscore), Precision (\mpscore), and Recall (\mrscore) in Tables  \ref{tab:results_known_http_malware} and \ref{tab:results_unseen_http_malware}.

\begin{table*}[t!]
\begin{center}
\scriptsize
    \begin{tabular}{p{4cm}p{0.8cm}p{0.8cm}p{0.8cm}p{0.2cm}p{0.8cm}p{0.8cm}p{0.8cm}p{0.2cm}p{0.8cm}p{0.8cm}} 
      \toprule 
      & \multicolumn{3}{c}{\textit{Macro}}&& \multicolumn{3}{c}{\textit{Weighted}}\\
    \cmidrule{2-4} \cmidrule{6-8}

         Classifier Name	&  \mpscore & \mrscore  & \mfscore &	& \wpscore & \wrscore  & \wfscore& &$FPR$ & $A$\\
    \midrule 
    \midrule 
          
          \multicolumn{10}{c}{\textbf{I. Dataset: APTs vs. Legitimate}}\\
         \earlycrow\	&94.48&	91.67&	93.02&&	98.07&	98.11&	98.08&&	0.74&	98.11	  \\
         Baseline	& \textbf{98.04}&	75.00&	82.33&&	96.37&	96.22&	95.63&&	\textbf{0.00}&	96.22	  \\
         \earlycrow-HTTPS	&94.68&	\textbf{92.81}&	\textbf{93.72}&&	\textbf{98.25}&	\textbf{98.28}&	\textbf{98.26}&&	0.74&	\textbf{98.28}  \\
         Baseline-HTTPS	&96.70&	56.82&	60.29&&	93.90&	93.47&	91.10&&	\textbf{0.00}&	93.47 \\

          \multicolumn{10}{c}{\textbf{II. Dataset: Botnets vs. Legitimate }}\\   
         \earlycrow\	& \textbf{96.77}&	\textbf{92.01}&	\textbf{94.25}&&	\textbf{99.25}&	\textbf{99.26}&	\textbf{99.24}&&	0.19&	\textbf{99.26}	  \\
         Baseline	&95.08&	78.85&	85.06&&	98.26&	98.35&	98.16&&	0.19&	98.35  \\
         \earlycrow-HTTPS	& 95.49&	81.48&	87.12&&	98.46&	98.53&	98.40&&	0.19&	98.53  \\
         Baseline-HTTPS	& 48.25&	50.00&	49.11&&	93.14&	96.51&	94.79&&	\textbf{0.00}&	96.51	  \\

        \multicolumn{10}{c}{\textbf{III. Malicious vs. Legitimate}}\\   
         \earlycrow\	& 94.77&	\textbf{91.60}&	\textbf{93.11}&&	\textbf{97.45}&	\textbf{97.51}&	\textbf{97.46}&&	0.93&	\textbf{97.51}  \\
         Baseline	& \textbf{95.89}&	76.10&	82.62&&	94.96&	94.85&	94.15&&	0.19&	94.85	  \\
         \earlycrow-HTTPS	& 94.27&	89.22&	91.54&&	96.92&	97.01&	96.92&&	0.93&	97.01	  \\
         Baseline-HTTPS	& 95.22&	54.76&	56.18&&	91.44&	90.53&	86.86&&	\textbf{0.00}&	90.53	  \\

  \bottomrule\\\\[-10pt] 
    \end{tabular}
     \end{center}

   \raggedright
   \caption{Unseen malware classification performance.}\label{tab:results_unseen_http_malware}
\end{table*}

\subsection{Known Malware Classification Performance} \label{sec:EarlyCrowevaluation_known}

In this experiment, we randomly split the training and testing sets ten times. Then, we take the average performance under two constraints. First, the malware should be presented in both sets. Second, the infected hosts and the destination \cc\ server should be unique and not leaked from training to testing.  

\earlycrow\ obtains the best performance with \mfscore\ of 93.89\%, 95.9\%, and 95.06\% for the three datasets. 
The performance of \earlycrow\ continues to achieve higher than baseline for \mpscore\ and \mrscore. 
We also note that the baseline achieves almost similar for APT and botnet (-0.21 $\Delta$\mfscore), while \earlycrow\ achieves better against the botnet (+2.01 $\Delta$\mfscore). This confirms that improving APT detection can help detect another type of malware, such as botnets. 
The FPR of \earlycrow\ are minimised to 0.40\%, 0.48\% and 0.86\%, which is also lower than the baseline.
Therefore, The detection performance on known malware for \earlycrow\ is not at the expense of higher FPR, which is critical to observe for SOC analysts. 

Even in HTTPS mode, which cannot take advantage of plaintext HTTP features such as headers or URL details, \earlycrow\ still outperforms the baseline on both three tasks, scoring 92.79\%, 95.84\%, and 94.89\% of \mfscore, respectively. Analogous to HTTP mode, \earlycrow-HTTPS achieves better \mfscore\ against botnets compared to APT ($\Delta$ +3.05\%).
Note that \earlycrow\ can operate almost similarly on both modes on known malware. 
However, we observe the $\Delta$\mfscore\ of the baseline between the two modes for the three tasks are 17.27\%, 5.45\%, and 7.21\%, which is a significant loss compared to \earlycrow, which are only 1.1\%, 0.06\%, and 0.17\%. 
However, we will investigate that on unseen malware in the next section. 

%


\subsection{Unseen Malware Classification Performance}  \label{sec:EarlyCrowevaluation_unseen}
For this experiment, we train our classifiers on the training set used for our measurement study. Then we evaluate the performance against the unseen malware described in Table \ref{tab:EarlyCrowdataset}.
\earlycrow\ obtains the best performance with \mfscore\ of 93.02\%, 94.25\%, and 93.11\% for the three datasets. 
While the baseline can achieve better \mpscore\ and \mrscore\ in a single task, the other tasks $\Delta$'s range fluctuates from 23.04\% to 19.79\% compared to the range from 4.76\% to 3.17\% for \earlycrow. The primary reason for the considerable gap is the baseline struggles to detect unseen malware samples. For example, \mrscore\ against APTs is 75\%, while \earlycrow\ achieves 91.67\%.
%

%
%

On all three tasks in HTTPS mode, \earlycrow\ surpasses the baseline, achieving 93.72\%, 87.12\%, and 91.54\%. 
Note that \earlycrow\ can operate almost similarly on both modes on unseen malware for the first and third datasets. However, the \mfscore\ is decreased when detecting only botnets with a margin (-7.13\%) compared to a severe loss for the baseline (-35.95\%). 
Also, the $\Delta$ between the \mpscore\ and \mrscore\ is relatively stable for the APT (1.87\%). However, the gap is growing faster against botnets (14.01\%).
The same conclusion can be held for the baseline, where the difference is even larger (39.88\%). 
We discuss the importance of using HTTP features for APT, the chances to improve the detection against botnet in HTTPS mode, and how that affects the FPR performance in Section \ref{sec:EarlyCrowDiscussion}.

\subsection{Features Diversity} 
\begin{figure}[h]
\centering
\includegraphics[width=8cm,height=6cm]{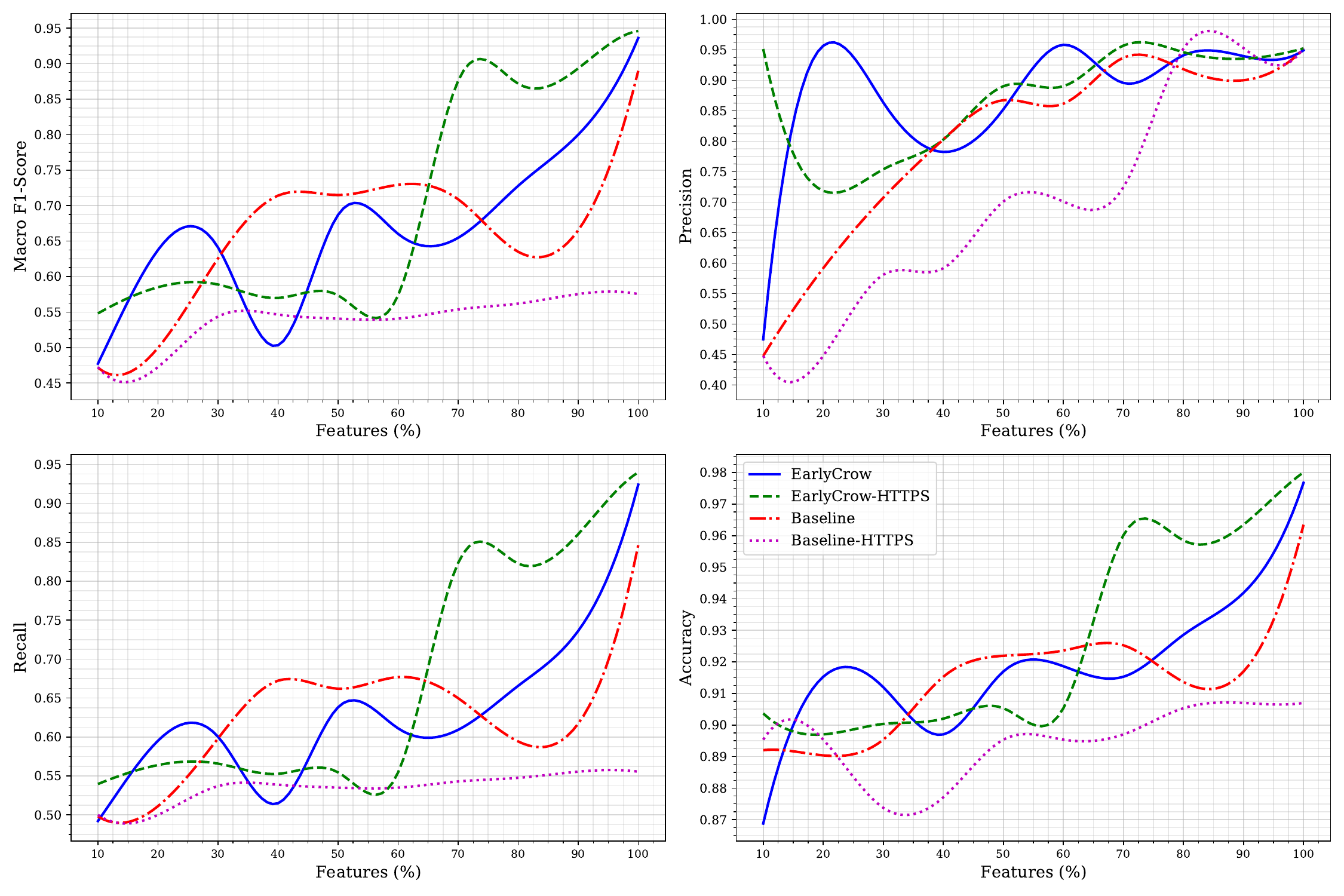}
\caption{Effect of using only the top \% of features.}\label{fig:feautres_metrics}
\end{figure}

Detecting APTs necessitates a spread of features, as presented in Section \ref{sec:EarlyCrow_threat_model}. 
In Figure \ref{fig:feautres_metrics}, we show the extent to which additional features affect the performance of the various classifiers based on the second experiment (Section~\ref{sec:EarlyCrowevaluation_unseen}).
The first 10\% of features for \earlycrow\ show rapid improvements in terms of precision but with poor recall.
Also, stronger features between 48\% and 62\%  can improve the performance of \mfscore\ up to 92.26\% for \earlycrow-HTTPS.
Adding more features afterwards increases the detection rate, enabling more unseen APTs to be detected. 
Furthermore, a detection system with diverse and strong features will require more time and resources to evade as opposed to one that relies on a few particular features \cite{wang2021crafting}. 

\subsection{Feature Importance} 
\begin{figure*}[!t]
\centering
\includegraphics[width=16cm,height=13cm]{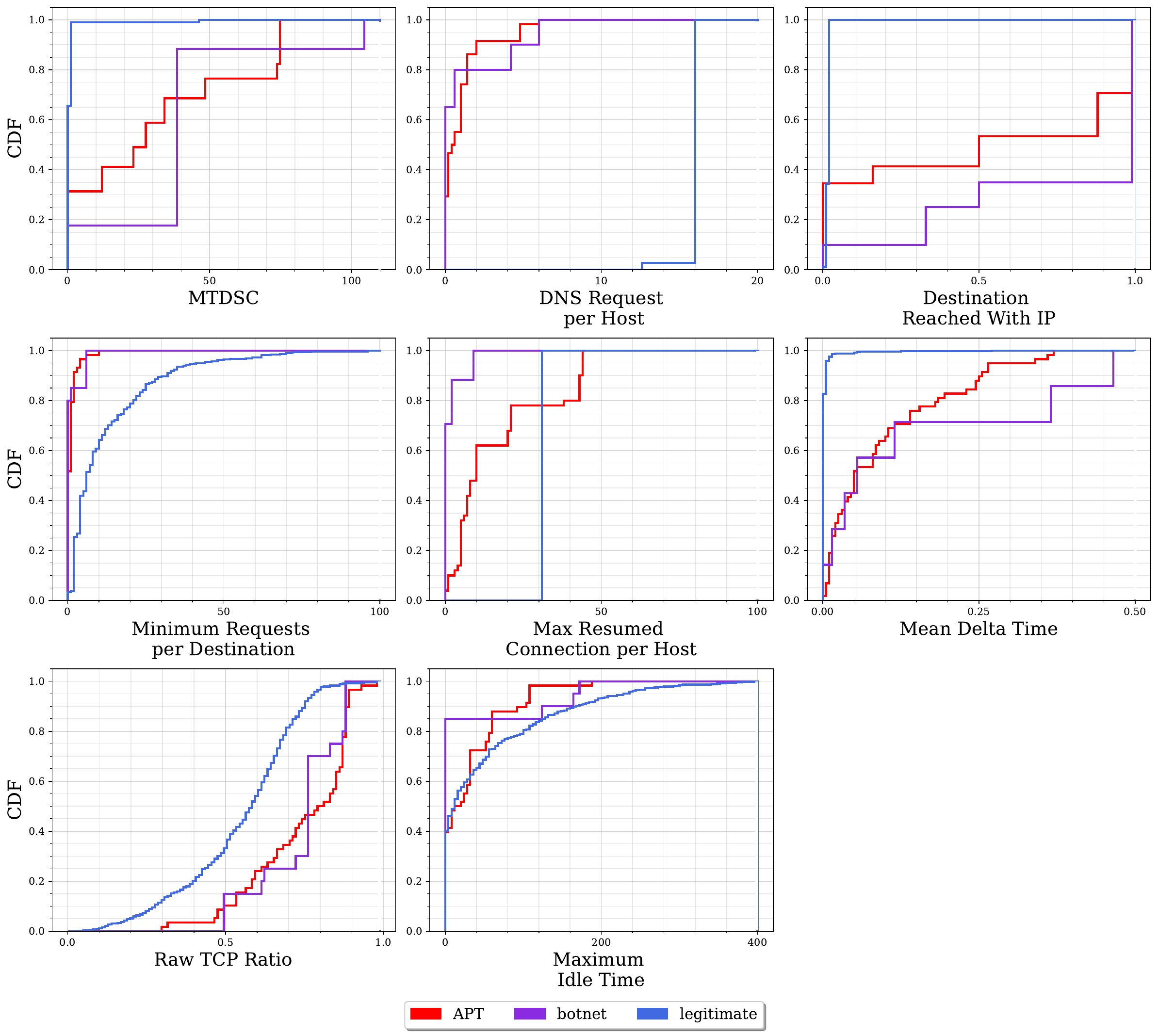}
\caption{Cumulative distribution of top features gains on the testing set for} \earlycrow-HTTPS. \label{fig:feature_importance}
\end{figure*}

We investigate the feature importance of the experiment of the third dataset in HTTPS mode, which comprises APTs, botnets, and legitimate samples, because it is the one closest to a realistic scenario for APT hunting. 
Figure \ref{fig:feature_importance} illustrates the top features based on their information gain. 
%
MTDSC is an effective feature that shows that 82\% of hosts infected with APTs and botnets spend up to 73.7 sec and 38.5 sec, which are higher than those for legitimate connections that spend up to 1.1 sec, confirming the typical HTTP browsing behaviour. 
The longer time for APTs indicates the use of \textit{fallback channel}, which is generally established after a long time.
Next, 90\% of DNS requests for host per connection are lower in APTs than botnets and legitimate with 2, 6, and 19, respectively. 
Interestingly, hosts infected with APTs have higher connections reaching further destinations by IP without \textit{domain resolution} as \textit{fallback channels}. This is consistent with our measurement study in Section \ref{sec:measurements}. 
70\% of APTs use such an approach with 88\% or less for their connections while only 1\% for the legitimate.

While 60\% of legitimate connections are resumed six times or less, botnets and APTs are rarely disconnected, with two-thirds lower. This confirms the malicious connections are weakly imitating legitimate behaviour (\textit{web protocol TTPs}).
Next, the mean delta interarrival time between packets during a \pairflow\ shows that APTs and botnets are significantly slower than legitimate, i.e., at 95\%, the mean delta time at most $33.5\times 10^{-2}$, $46\times 10^{-2}$  and $ 0.5\times 10^{-2}$ sec, respectively.
We confirm that APTs tend to switch from HTTP to raw TCP for malicious operations (\textit{non-application protocols}). Within a \pairflow, we find 50\% of APTs rely on 81.09\% (58.35\% for legitimate) of the whole exchange packets on raw TCP, indicating the adversary uses HTTP as camouflage while still relying on TCP for many tasks.
Nonetheless, the APTs and botnets are faster regarding the difference between data packets. 
APTs and botnets tend to be shorter/faster than legitimate, where 90\% of them take 104, 124, and 168 sec, respectively.

\subsection{Attacks Investigation} 
In Table \ref{tab:results_detect_rate}, we breakdown the results of \earlycrow-HTTPS on unseen malware. 92\% of unseen malware are detected with at least one \cc\ communication, and 64\% of different malware are fully detected. However, one server belonging to StrongPity is not detected at HTTPS. We found that StrongPity does not use a fallback channel, and its measurement reflects a legitimate one. At HTTP mode, \earlycrow\ managed to detect StrongPity because of its malicious URL characteristics, such as using \texttt{.exe} file extension and lack of rich web server (i.e., No. of URLs distinct) compared to the data exchanged. Also, some \cc\ servers belonging to OnionDuke and Zeus manage to evade detection. These servers are established as fallback channels with minimum data transfers, evading many features. Since the malware is detected on a specific machine, we recommend a SOC analysis to sanitize the victim machine from the malware to stop other possible \cc\ communications.  

%
%
%
%
%

%
\subsection{Features and TTPs Correlation} \label{sec:featuresTTPsCorr}

\begin{figure*}[!t]
\centering
\includegraphics[width=16cm,height=6cm]{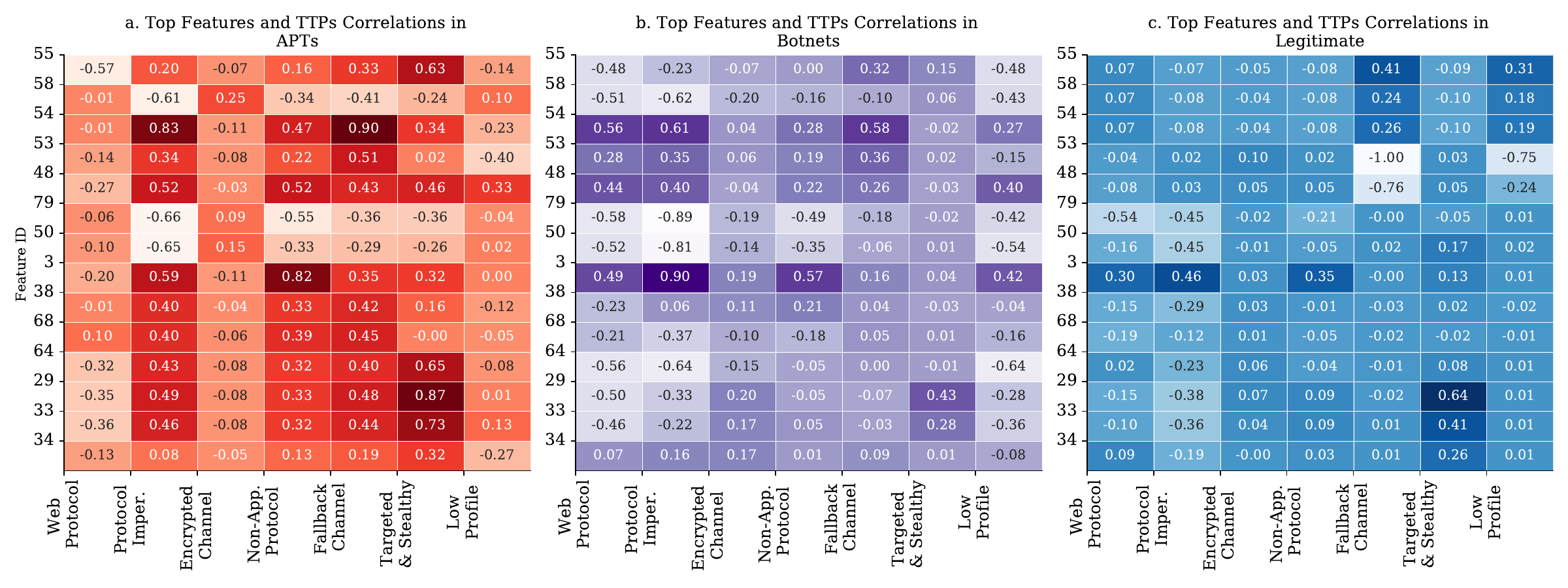}
\caption{Heatmap for \earlycrow-HTTPS.}\label{fig:heat_mao}
\end{figure*}

In Figures \ref{fig:heat_mao}.a-c, we present a heatmap on the third dataset based on the unseen malware classification experiment (Section~\ref{sec:EarlyCrowevaluation_unseen}). We confirm that legitimate traffic uses some APT TTPs, increasing the overlap, which confirms our findings of the measurement study. With the help of other features, detecting the malicious one can be accomplished. 
For example, many legitimate connections can use \textit{encrypted channel}. 
It has a major overlap between all classes; therefore, \earlycrow\ is not able to detect the malicious use of such TTP. However, the encrypted channel is ineffective in hiding other TTPs against \earlycrow\ because several of our features do not rely on deciphering the HTTPs payload. 
We find that features (\#33, \#64, \#68) have a solid signal to separate APTs from botnets, but not legitimate ones. 
The number of connected hosts to a destination shows the APT mimic the legitimate, unlike botnets, where the infection spreads throughout several hosts. 
Feature (\#58) does not show correlation to \textit{web protocol} in APTs and legitimate due to the standard use of DNS requests ratio to the response. 
Still, botnet shows a higher negative correlation due to DGA use. 
We also identify minor overlaps in further TTPs. 
For instance, \textit{Targeted and stealthy} TTP has a small difference in linear correlation between APTs and legitimate (+0.73, +0.41). 
Both APTs and legitimate behave similarly in terms of points below the average of SMA; however, the linear relationship is slightly larger in APTs than in legitimate (+0.87, +0.64). 

\begin{table}[h]
\scriptsize
\centering
    \begin{tabular}{lcc||lcc}
    
      \toprule 

        \textbf{Malware}& \makecell{\multirow{2}{*}{\textbf{\makecell{\cc \\ Servers}}}} &\makecell{\multirow{2}{*}{\textbf{\makecell{Detection \\ (\%)}}}} & \textbf{Malware}& \makecell{\multirow{2}{*}{\textbf{\makecell{\cc \\ Servers}}}}&\makecell{\multirow{2}{*}{\textbf{\makecell{Detection \\ (\%)}}}}  \\&&&&&\\
    \midrule 
    \midrule 
Ammyy & 8 & 100& ImminentMonitor & 4 & 75  \\
ChChes & 1 & 100 & Magic-Hound & 3 & 100   \\
CobaltStrike & 2 & 100& OnionDuke& 6 & 33.34 \\
Dridex & 2 & 100  & PoisonIvy& 1 & 100\\
Emotet & 13 & 53.84 &  Ramnit & 2 & 100\\
Empire & 5 & 100 & StrongPity &1 & 0 \\
FlawedAmmy& 4 & 100 & Zeus & 3 & 33.34 \\

  \bottomrule\\\\[-10pt] 
    \end{tabular}
   \caption{Detection rate on unseen malware over HTTPS.}\label{tab:results_detect_rate}
\end{table}

On the other hand, other features are powerful in identifying a separable correlation of malicious use of TTPs between classes. 
We find feature (\#55) and  \textit{web protocol} are negatively correlated with APTs and botnets but almost no correlation with legitimate. 
Feature (\#54) can identify the \textit{protocol impersonation} in APTs and botnets while the legitimate seems to not correlate with it. 
Non-Application Protocol TTP has a large positive correlation to APTs (+0.82), while the legitimate has weaker ones (+0.35) due to the common use of frequent Raw TCP in APTs. 
Finally, it should be known that none of these correlations is used for detection due to some of the possible spurious correlations. 
For instance, a spurious correlation can be found between the feature (\#55) and \textit{fallback channel}. 
Thus, we use it for analysis only to explain the differences in relationships between features and TTPs across all classes.


\subsection{Discussion}\label{sec:EarlyCrowDiscussion}

As presented previously, this section discusses the performance of all classifiers and highlights the difference between known (Section \ref{sec:EarlyCrowevaluation_known}) and unseen malware (Section \ref{sec:EarlyCrowevaluation_unseen}). 
%
%
%
For \earlycrow, the performance loss ($\Delta$\mfscore) between known and unseen is marginally low (1.97\%) in the third dataset, while the baseline suffers a loss of 9.6\%.
Likewise, the performance loss ($\Delta$\mfscore) for \earlycrow\ between the two modes of known malware compared to unknown malware is +27.27\% from 1.1\% to 1.4\%. For the baseline, the $\Delta$ is dramatically increasing by up to +108.16\%, namely, from 17.27\% to 35.95\%.

%

As pointed out in Section~\ref{sec:EarlyCrowevaluation_unseen}, we observe a high difference between \mpscore\ and \mrscore\ against unseen botnets, which is not the case for unseen APTs. We conclude that the plaintext HTTP features play a center role in detecting unseen botnets.  
This is also the case for both APT and botnet detection for FPR reduction, where these features can reduce the FPR by nearly \%50 for \earlycrow\ and the baseline. 
That suggests more work on HTTPS mode has a high chance to improve the stability on differences between \mpscore\ and \mrscore, in addition to the FPR reduction close to \earlycrow-HTTP. 
%
%
%
%
We suggest investigating if using time series techniques based, such as DSP, on the interarrival time of packets as a time series problem can improve the HTTPS-mode classifier. Using suggested techniques can also help cover Cases III and IV discussed in Section~\ref{sec:EarlyCrow_threat_model}. We will investigate such approaches in the future work.

\subsection{Limitations}
In the early stages of an infection, it is difficult to tell how many suspicious activities are linked to more advanced attacks and how many are mainstream malware variants because \earlycrow\ is geared toward detecting the first stages of infection. 
We recommend tracking the malware activities over a longer period of time with various APT campaign usage. 
This could be done by placing \earlycrow\ in a targeted entity such as a sovereign entity or large financial institution network over months and reevaluating \earlycrow\ reports by security experts for further improvement.
%
%

\section{Related Work}\label{sec:earlycrow-related-work}

There is very limited previous work on detecting APTs at the network level. 
Detecting \cc\ in general is the closest area.
In our approach we test several features from the literature which can be relevant for APTs, including URLs and UA features \cite{oprea2018made,oprea2015detection,invernizzi2014nazca,perdisci2010behavioral}, traffic exchange bytes \cite{bartos2016optimized,bilge2012disclosure,tegeler2012botfinder, oprea2018made}, HTTP content types \cite{oprea2018made,oprea2015detection}, and GET and POST ratio \cite{perdisci2010behavioral,oprea2018made}. Besides directly using such features, \earlycrow\ pivots them into host, destination and URL profiles, and combines them in contextual summaries. 

Some previous works focus on detecting APTs in addition to other kinds of malicious communications \cite{oprea2015detection,oprea2018made}. 
Oprea et al. \cite{oprea2015detection} propose a belief propagation (BP) algorithm to detect early-stage infection of APTs. 
They model enterprise communication using a bipartite graph with two vertices, hosts, and domains based on simulated attacks.
Once the detector identifies a malicious remote host or domain based on several features, BP identifies communities of malicious domains with similar features that are part of the same attack campaign. 
Domain scores are calculated as a supervised linear regression weighted sum of features.
As discussed in Section~\ref{sec:EarlyCrowevaluation}, APTs tend to infect a lower number of hosts than botnets. 
Therefore, \earlycrow\ considers other features based on different TTPs discussed in Section \ref{sec:EarlyCrow_threat_model}. 
\earlycrow\ is closer to \made~\cite{oprea2018made}, which instead uses web proxy logs at the edge of an enterprise network to detect malicious \cc\ communications, including APTs. 
\made\ leverages features related to the communication, HTTP request, response and its content, URL, and UAs.
These are used by a random forest classifier to assign a risk score for each connection. 
As discussed in Section \ref{sec:EarlyCrowevaluation}, \made\ is not as effective on HTTPS traffic, which is nowadays harder to intercept and decrypt due to technical and legal requirements. 
In addition, \earlycrow\ considers five other TTPs besides the \textit{Web Application Protocol} TPP at the heart of \made.

ExeceScent~\cite{nelms2013execscent} detects \cc\ domains by clustering incoming requests into five templates, including median URL path, URL query component, User-Agent, other headers, and destination network. 
These templates are used to estimate similarity scores to predefined Control Protocol Templates (CPT) centroids. 
However, this is open to evasion if an adversary copies the UA of the victim machine from the Windows Registry \cite{bortolameotti2017decanter}. 
In addition, it is not possible to extract most HTTP header features when HTTPS is in use, which hinders the generalization process and may result in mixing APTs with legitimate in many clusters. 
A related approach~\cite{bartos2016optimized} adopts similar features, only using histogram bins which also can be evaded using HTTPS. 
BAYWATCH \cite{hu2016baywatch} is a filtering system to detect the beaconing of infected hosts. 
Universal and local whitelists are filtered, and then beaconing can be detected using the Discrete Fourier Transform (DFT) and Gaussian Mixture Model (GMM), awarding a high Agglomerative Hierarchical Clustering (ACF) score for strong periodicity.
BAYWATCH filters URLs and domains that are likely to be legitimate. 
Unprocessed connections with all previous features are sent to a random forest for classification.
BAYWATCH can be computationally expensive for only beaconing behaviour, and many APTs also have non-beaconing connections. 
\earlycrow\ detects malicious connections regardless of their pattern.
Finally, Kitsune \cite{mirsky2018kitsune} adopts an ensemble of autoencoders, proving the efficiency of unsupervised deep learning to detect classic attacks such as ARP poisoning and SYN DoS, which are rarely used by APTs. 
In this paper, we avoid using deep learning because of the scarce dataset representing various APTs TTPs, which is essential for deep learning models.


\section{Conclusions}

In this paper, we presented a threat model for APTs informed on the TTPs used by adversaries to avoid existing NIDS. 
Informed by that, we designed and implemented \earlycrow, a random-forest-based classifier which can detect APT malware network activities that are missed by currently deployed defence mechanisms.
We recommend using \earlycrow\ as an additional layer of defence, besides SIEM, Host Intrusion detectors (HIDS), and domain detectors.
Building \pairflow\ while considering an APT-specific threat model enabled \earlycrow\ to catch TTPs used in APTs with promising performance. 
However, \pairflow\ can be used for further tasks beyond malicious HTTP(S) and domain detection, including non-application protocols (Raw TCP, UDP, ICMP, and Socket Secure (SOCKS)). Adversarial attacks against \pairflow\ fields and \earlycrow\ features, their robustness, and deployment issues are left to future work.

\bibliographystyle{ieeetr.bst}

\end{document}